\documentclass{article}

\usepackage{authblk}           
\usepackage[a4paper]{geometry} 
\usepackage{url}               
\usepackage{graphicx}          
\usepackage{grffile}           
\usepackage{booktabs}          
\usepackage{amsmath}           
\usepackage{caption}           
\usepackage{rotating}          
\usepackage{float}			   
\usepackage{hyperref}          
\usepackage{pdflscape}         
\usepackage{longtable}         
\usepackage{comment}

\title{Multi-indication evidence synthesis in oncology health technology assessment}

\author[1]{Janharpreet Singh}
\author[3]{Sumayya Anwer}
\author[2]{Stephen Palmer}
\author[2]{Pedro Saramago}
\author[4]{Anne Thomas}
\author[3]{Sofia Dias}
\author[2]{Marta Soares}
\author[1]{Sylwia Bujkiewicz}

\affil[1]{Biostatistics Research Group, Department of Population Health Sciences, University of Leicester, Leicester, UK}
\affil[2]{Centre for Health Economics, University of York, York, UK}
\affil[3]{Centre for Reviews and Dissemination, University of York, York, UK}
\affil[4]{Leicester Cancer Research Centre, University of Leicester, Leicester, UK}

\begin{document}

	\maketitle

	\begin{abstract}
		\textbf{Background}:
		Multi-indication cancer drugs receive licensing extensions to include additional indications, as trial evidence on treatment effectiveness accumulates. We investigate how sharing information across indications can strengthen the inferences supporting Health Technology Assessment (HTA).
		
		\textbf{Methods}:
		We applied meta-analytic methods to randomised trial data on bevacizumab, to share information across oncology indications on the treatment effect on overall survival (OS) or progression-free survival (PFS), and on the surrogate relationship between effects on PFS and OS. 
            Common or random indication-level parameters were used to facilitate information sharing and the further flexibility of mixture models was also explored.

		\textbf{Results}:
 Treatment effects on OS lacked precision when pooling data available at present-day within each indication separately, particularly for indications with few trials. There was  no suggestion of heterogeneity across indications. 
  
Sharing information across indications provided more precise inferences on treatment effects, and on surrogacy parameters, with the strength of sharing depending on the model. When a surrogate relationship was used to predict OS treatment effects, uncertainty was only reduced with sharing imposed on PFS effects in addition to surrogacy parameters. 

Corresponding analyses using the earlier, sparser (within and across indications) evidence available for particular HTAs showed that sharing on both surrogacy and PFS effects did not notably reduce uncertainty in OS predictions. Limited heterogeneity across indications meant that the added flexibility of mixture models was unnecessary. 
		
		\textbf{Conclusions}:
		Meta-analysis methods can be usefully applied to share information on treatment effectiveness across indications in an HTA context, to increase the precision of target indication estimates. Sharing on surrogate relationships requires caution, as  meaningful precision gains will likely require  larger bodies of evidence and clear support for surrogacy from other indications. 	
	\end{abstract}
	
 \clearpage

	\section{Introduction}
	
		In health technology assessment (HTA) of cancer drugs, evidence on a relative treatment effect on overall survival (OS) supports assessments of the clinical and economic value of a drug, relative to a comparator, for a patient population defined by a particular indication (e.g., tumour type and stage of disease). 
		As HTA agencies seek to make a reimbursement decision soon after a drug is licensed, to ensure timely patient access to treatment, the evidence on treatment effectiveness may be limited. Regulatory agencies increasingly make accelerated licensing decisions based on a single trial, and often a treatment effect measured on a surrogate endpoint; for example progression-free survival (PFS) as a surrogate endpoint to OS \cite{DelPaggio2021}. 
		This can result in large uncertainty in effect estimates on OS, or effect estimates available for PFS alone, at the time the HTA is conducted.
		
		Some cancer drugs are, however, trialled across multiple indications over time. When the results in a new indication are reported, the drug's license may be extended to this new indication \cite{Garcia2020}.  Despite this accumulation of evidence over time, HTAs typically restrict their scope to consider evidence within a single target indication. Alternatively, sharing information across indications could strengthen the evidence base supporting decision-making on a new indication.
		The National Institute for Health and Care Excellence (NICE) in England and Wales has identified the development of evidence synthesis methods for multi-indication HTA of cancer drugs as a key priority \cite{NICE2021}.
		
		Explicit evidence synthesis models exist that share information across evidence sets \cite{Nikolaidis2021} by defining relationships between parameters corresponding to the different sets of evidence. 
            These relationships can be functional, hierarchical, multivariate, or based on a prior distribution in a Bayesian framework.
		For example, network meta-analysis, which allows for sharing of information across treatment comparisons, has become an established method to inform treatment effect estimates in HTA \cite{Dias2018}.
		Similarly, multivariate meta-analysis for sharing information across multiple outcomes, has been proposed to help validate a supposed surrogate endpoint as a good predictor of clinical benefit by modelling a surrogate relationship between the treatment effects on a surrogate endpoint (e.g PFS) and a final clinical outcome (e.g OS) \cite{Bujkiewicz2019a, Daniels1997, NICE2021}.

            Sharing of information has been previously applied in the multi-indication context. 
            Panoramic meta-analysis \cite{Hemming2012, Chen2014}, and models for the analysis of basket trials of histology independent oncology drugs \cite{murphy2021exploring, murphy2021modelling}, have used hierarchical models assuming full exchangeability (similarity) across indications where indication-level parameters vary according to a common distribution. 
            In these models, the data determine the strength of sharing. However, more flexible models may be required to support decision making. For example, models that do not presume full exchangeability, and models which allow clinical opinion to regulate the degree of sharing where there is uncertainty regarding the plausibility of sharing across particular indications due to clinical factors.

        Mixture models implement a sharing component and a component assuming independence, with the contribution of each component being regulated by a  probability parameter. This probability can either be pre-specified, estimated from the data, or informed by a prior distribution to incorporate external information.  
		The simplest mixture model consists of a weighted mixture of two components: an informative prior distribution (sharing component), and  a vague prior distribution \cite{Schmidli2014}.  
            Extensions include a model
            where the sharing component consists of a common parameter across strata \cite{Rover2019}, and a model where the sharing component consists of a hierarchical relationship across strata \cite{Neuenschwander2016}. 
            The latter model has been termed a partial exchangeability model and was originally proposed for the analysis of indication strata within an individual study.

            Mixture models have been applied in other contexts of evidence synthesis, but have not been formally evaluated for use in the multi-indication context. Notably, Papanikos \textit{et al} shared information on surrogate relationships (between treatment effects on surrogate and final outcomes) across treatment classes by assuming either full, or partial, exchangeability for the class-level surrogacy parameters \cite{Papanikos2020}. 
            They found that the models for sharing information can provide a reduction in uncertainty compared to a subgroup analysis, particularly where data are limited within classes, and that the partial exchangeability model can identify differences in surrogate relationships between classes to regulate the level of sharing.
		
           Applying mixture models to perform a multi-indication meta-analysis could allow sharing of information across similar indications whilst avoiding sharing from extreme indications, providing more robust inference in cases where the plausibility of full exchangeability across indications is uncertain. 
            These models could also allow for the inclusion of external information on the mixture probability representing the plausibility of sharing across indications. Furthermore, such models can be implemented to share  either on treatment effects on OS as a single final endpoint, or on the surrogate relationships between treatment effects on PFS and OS to predict an effect on OS for a particular indication.

		In this paper, we demonstrate the application of mixture models to multi-indication meta-analysis in oncology, to enable sharing  on individual endpoints and on a surrogate relationship, where a predicted treatment effect on OS is the key estimate of interest. We use the example of bevacizumab, one of the earliest multi-indication oncology drugs, which has been trialled across multiple cancer indications and been the subject of several NICE  technology appraisals (TAs).
		
		In the remainder of this paper, we describe the data set for the bevacizumab case study (Section \ref{sec:caseStudy}),  introduce the meta-analysis models implemented to share information across indications (Section \ref{sec:methods}),  present the results from applying the models to the case study data (Section \ref{sec:results}) and discuss our findings  (Section \ref{sec:discussion}).
		
	\section{Case study: bevacizumab} \label{sec:caseStudy}
		
			Bevacizumab is an angiogenesis inhibitor which is used as a targeted therapy for different solid tumours and has become the standard-of-care treatment for several advanced cancers \cite{Garcia2020}.
			It has received multiple licensing extensions from the European Medicines Agency, such that it is currently approved as a treatment for seven cancer indications. In the UK, it has been the subject of 11 NICE TAs across seven cancer indications.
			
			The evidence base for bevacizumab was established by performing a literature search to identify randomised controlled trial (RCT) evidence across its licensed indications in the advanced or metastatic cancer setting: breast cancer (BC), cervical cancer (CC), colorectal cancer (CRC), glioblastoma (GBM), non-small cell lung cancer (NSCLC), ovarian fallopian tube, and primary peritoneal (OFTPP) cancer, and renal cell carcinoma (RCC). 
            Evidence sources included NICE TAs, Cochrane Reviews, and the \textit{ClinicalTrials.gov} database.
			The scope was restricted to phase II/III trials where the treatment effect of bevacizumab as a monotherapy could be isolated.
			
\begin{sidewaysfigure}[ht]
				\centering
				\includegraphics[width=\textwidth]{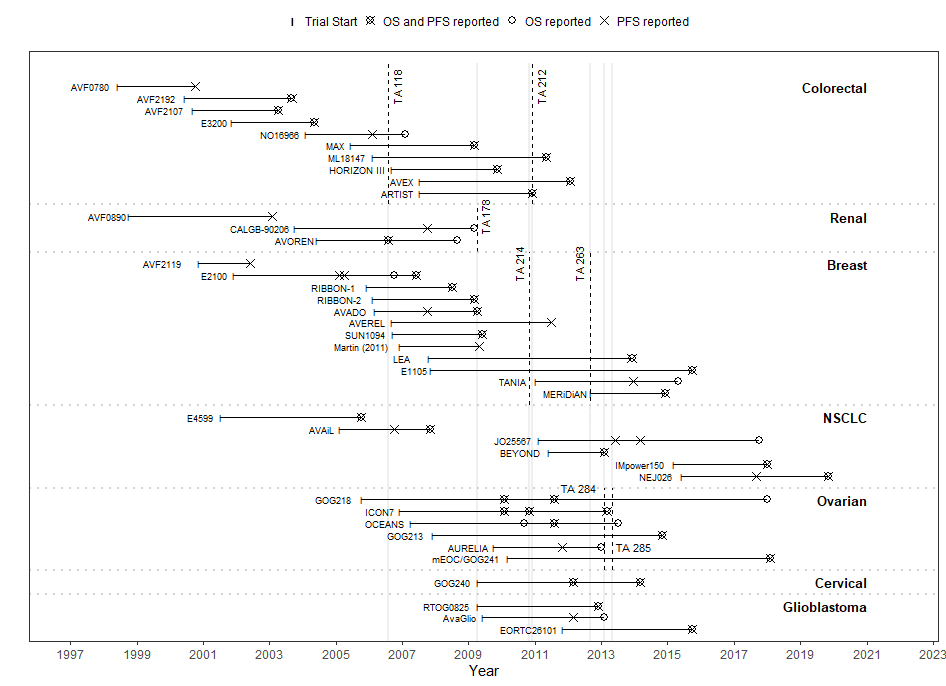}
				\caption{Plot summarises the accumulation of data on overall survival (OS) (circles) and progression-free survival (PFS) (crosses) from randomised controlled trials assessing bevacizumab within each cancer indication (labelled on far right) between 1997 and 2023. Vertical dashed lines indicate time points corresponding to particular technology appraisals (TAs) by the National Institute for Health and Care Excellence. NSCLC - Non-small cell lung cancer.}
				\label{fig:taTimelineSummary}
			\end{sidewaysfigure}

			Figure \ref{fig:taTimelineSummary} shows the development of RCT evidence on the relative effectiveness of bevacizumab from 1997 to 2023, and the time point(s) at which these trials reported results. 
			A total of 41 RCTs across seven cancer indications were identified. Most of these trials compared the addition of bevacizumab to backbone therapy, typically chemotherapy.  All 41 trials reported effect estimates for PFS, and 36 reported effect estimates for OS.
   
 Over the time period considered, bevacizumab was the subject of seven NICE TAs across four cancer indications. The appraisal time points are indicated in Figure \ref{fig:taTimelineSummary} with vertical dashed lines.  For our analyses, we will focus on the evidence available at present day and at two appraisal time points, an earlier time point with limited evidence and a later time point with more abundant evidence. The first appraisal is TA178, appraising the use of bevacizumab in RCC \cite{NICETA178}. At the time of this appraisal, there were three completed trials in the target indication (three reporting effect estimates for PFS and two for OS), and an additional 14 trials across three other indications (14 reporting effect estimates for PFS and  11 for OS).  The second appraisal is TA285 in ovarian cancer \cite{NICETA285}. By the time of this appraisal, there were  four trials in this indication (all reporting effect estimates for both PFS and OS), and an additional 28 trials across six other indications (28 reporting effect estimates for PFS, and  23 for OS). 
   We focus on these time points to demonstrate the methods in contrasting scenarios in terms of the availability of evidence.

			\clearpage
			
\begin{sidewaysfigure}[ht]
				\centering
				\includegraphics[width=\textwidth]{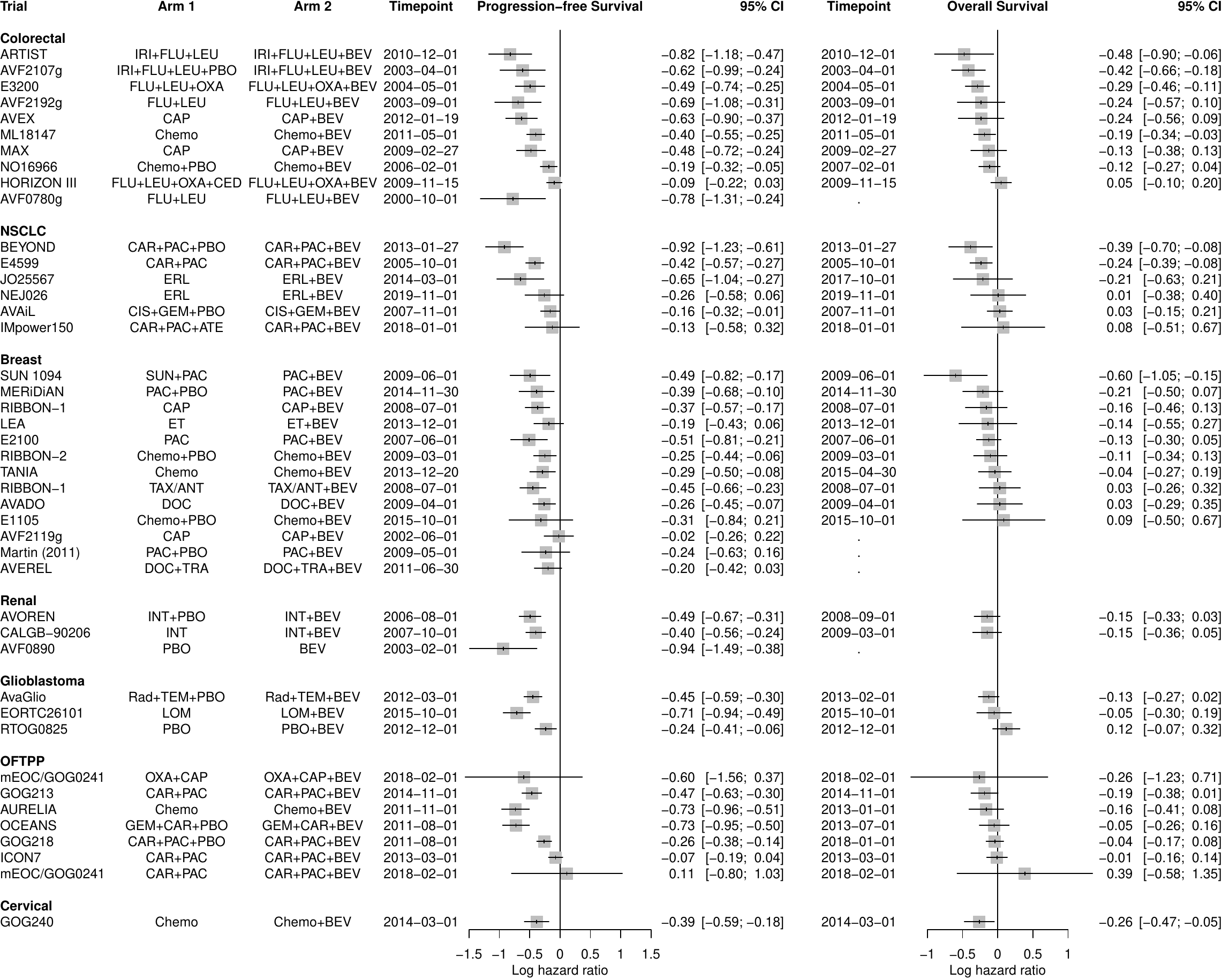}
				\caption{Forest plot summarising log-transformed hazard ratio estimates on progression-free survival (PFS) and overall survival (OS) from randomised controlled trials assessing bevacizumab across cancer indications. Estimates are ordered by OS effect size within each cancer indication. BEV - bevacizumab. NSCLC - Non-small cell lung cancer. OFTPP - Ovarian, fallopian tube and primary peritoneal. A list of full treatment names is provided in Appendix \ref{sec:treatmentNames}.}
				\label{fig:observedPresentDay}
			\end{sidewaysfigure}

			Figure \ref{fig:observedPresentDay} presents the final log hazard ratio (HR) estimates, on PFS (left) and OS (right), reported by trials across all cancer indications.
			The estimates are ordered by OS effect size (largest to smallest) within each indication.
			The majority of trials show a significantly favourable effect of bevacizumab on PFS and this is consistent both within and across indications.
			There is more uncertainty around effect estimates on OS, where point estimates indicate a favourable effect of bevacizumab but statistical significance is achieved in only seven trials.
                For the indications with relatively more trials (CRC, NSCLC and OFTPP), there is evidence of between-studies heterogeneity in PFS estimates (demonstrated by non-overlapping confidence intervals), although this is more difficult to establish on OS.
			The data show that larger effect estimates on PFS correspond to larger effect estimates on OS, particularly for CRC, NSCLC, and OFTPP cancer, which is suggestive of a potential surrogate relationship between these outcomes.

			\clearpage
		
	\section{Methods} \label{sec:methods}
In this section, we define the models used to perform the multi-indication meta-analysis (\ref{sec:modelDescription}) and detail their implementation (\ref{sec:modelImplementation}). All models are defined in a Bayesian framework and are implemented using Markov chain Monte Carlo (MCMC) simulation. 
		We describe the models in two parts. The first part relates to the within-indication model components (Section \ref{sec:likelihood}). In this part, we  synthesise treatment effects on OS (univariate models) or, alternatively,  on surrogate relationships between treatment effects on PFS and OS (bivariate models). Both approaches synthesise data on the log HR scale. The second part relates to the between-indications component of the models (Section \ref{sec:sharingModels}). 
We consider five alternative approaches to facilitate different levels of sharing of information across indications on the indication-specific parameters.  We apply each of the five sharing approaches to both the univariate and bivariate models.    
We also consider how to interpret parameters in the context of HTA decision-making and how to predict a treatment effect on OS, the key parameter of interest, using the bivariate models (Section \ref{sec:prediction}).
						
		\subsection{Models for multi-indication meta-analysis} \label{sec:modelDescription}
			
			\subsubsection{Within-indication component: within-study and between-studies models} \label{sec:likelihood}
				Modelling approaches considered at the within-indication level use standard meta-analytic specifications for the within-study (likelihood) and the between-studies models.
				
The univariate models use a standard contrast-based meta-analytic approach assuming the normal-normal hierarchical (random effects) model as described by Sutton and Abrams \cite{sutton2001bayesian}. This assumes that the relative treatment effects, $Y_{ij}$ (in  study $i$ within indication $j$), are normally-distributed, with associated standard errors $\sigma_{ij}^2$ (assumed known), and mean defined by the true treatment effects $\delta_{ij}$,
					\begin{equation}
						Y_{ij} \sim N(\delta_{ij}, \sigma_{ij}^2).
					\end{equation}
					At the between-studies level, the true treatment effects are assumed to be exchangeable across studies within each indication,
					\begin{equation}
						\delta_{ij} \sim N(d_j, \tau_j^2),
					\end{equation}
					where $d_j$ is the pooled effect for indication $j$ and $\tau_j$ is the associated (within-indication) between-studies standard deviation.  A weakly informative half-Normal prior distribution is placed on the between-studies standard deviation parameters, $\tau_j \sim |N(0, 0.5^2)|$, as recommended by Rover \textit{et al} \cite{Roever2021}, which are assumed to be independent across indications in all analyses. For the purpose of multi-indication meta-analysis, we seek to share information on the pooled effects, $d_j$.
		
The bivariate modelling approach uses the formulation by Daniels and Hughes \cite{Daniels1997} to examine a linear surrogate relationship between the treatment effects on the surrogate endpoint and on the final clinical outcome within each indication. We assume that, in each study $i$ and indication $j$, data on the effects are available (e.g., as log HR estimates) on a surrogate endpoint, $Y_{1ij}$, and a final outcome, $Y_{2ij}$, with associated standard errors $\sigma^2_{1ij}$ and $\sigma^2_{2ij}$, respectively, and within-study correlation $\rho_{wij}$.  At the within-study level, the correlated observed effects are assumed to follow a bivariate normal distribution,
					\begin{equation}
						\begin{aligned}
							&\begin{pmatrix}
							Y_{1ij} \\
							Y_{2ij}
							\end{pmatrix} \sim N \left( \begin{pmatrix}
							                            \delta_{1ij} \\
							                            \delta_{2ij}
							                            \end{pmatrix}, \begin{pmatrix}
							                            \sigma_{1ij}^2                   &\sigma_{1ij}\sigma_{2ij}\rho_{wij} \\
							                            \sigma_{1ij}\sigma_{2ij}\rho_{wij} &\sigma_{2ij}^2
							                            \end{pmatrix} \right) \\
						\end{aligned}
					\end{equation}
					where $\delta_{1ij}$ and $\delta_{2ij}$ are the correlated true effects on the surrogate endpoint and final clinical outcome, respectively.  The within-study correlation between the effects on the two outcomes is rarely reported, but can be informed by external data or assigned a prior distribution. We use a vague uniform prior distribution in our implementation $\rho_{wij} \sim U(-1, 1)$. 
     
     At the between-studies level, the true effects on the final outcome, $\delta_{2ij}$, are assumed to have a linear relationship with the true effects on the surrogate endpoint $\delta_{1ij}$,
					\begin{equation} \label{eqn:surrogacyBetweenStudies}
						\delta_{2ij} \sim N \left( \lambda_{0j} + \lambda_{1j} \delta_{1ij}, \psi^2_j \right)
					\end{equation}
					where $\lambda_{0j}$ and $\lambda_{1j}$ represent the intercept and slope respectively, and $\psi^2_j$ is the variance of the true effects on the final outcome conditional on the true effects on the surrogate endpoint (conditional variance) within indication $j$.	
     
     These parameters can be used to assess the strength of the surrogate relationship using the criteria proposed by Daniels and Hughes \cite{Daniels1997}; where a null effect on the surrogate endpoint should imply a null effect on the final outcome ($\lambda_{0j} = 0$), the slope should indicate an association between the endpoints ($\lambda_{1j} \neq 0$), and the conditional variance measures to what extent the treatment effect on the final outcome can be predicted from the effect on the surrogate endpoint (with $\psi_j^2 = 0$ corresponding to perfect predictions).
					Vague normal prior distributions are placed on the true effects corresponding to the surrogate endpoint, $\delta_{1ij} \sim N \left( 0, 10^2 \right)$, which are assumed to be independent across studies. 
                    For the purpose of multi-indication meta-analysis, we seek to share information on the intercept ($\lambda_{0j}$), slope ($\lambda_{1j}$), and conditional variance parameters ($\psi^2_j$), across indications.   
													
			\subsubsection{Between-indications component} \label{sec:sharingModels}
				
We apply five distinct between-indications models making different assumptions about the relationships between indication-specific parameters, and therefore, implying different degrees of sharing of information across indications.  To allow generalisability to both univariate and bivariate modelling approaches, we henceforth denote the indication-level parameters by the vector $\boldsymbol{\theta}_j$. Note that, in the bivariate approach, the following five models share information for each of the three surrogacy parameters (intercept, slope and conditional variance), but the sharing relationships are implemented independently for each parameter such that the level of sharing could differ between the parameters. 
				
				\paragraph{Independent parameters (IP) model}
					
					Here, independent prior distributions are placed on the $\boldsymbol{\theta}_j$  ,
					\begin{equation}
						\boldsymbol{\theta}_j \sim P().
					\end{equation}	
					When modelling data on a treatment effect (univariate model), we use a vague normal prior distribution for each indication, $P(d_j) = N(0, 10^2)$.
					When considering parameters of a surrogate relationship (bivariate model),  the vague prior distributions used for each of the parameters are; $P(\lambda_{0j}) = N(0, 10^2)$, $P(\lambda_{1j}) = N(0, 10^2)$ and $P(\psi_j) = |N(0, 0.5^2)|$.

				In the IP model, the indication-level parameters are informed by direct (within-indication) evidence only and there is no sharing of information across indications. This model provides a  reference to compare the treatment effect estimates from other models where sharing is allowed. 
    
				\paragraph{Common parameter (CP) model}
					
					Here, we assume a common overall treatment effect parameter, $\boldsymbol{\theta}$,  which pools the indication-specific parameter values,
					\begin{equation}
						\boldsymbol{\theta}_j = \boldsymbol{\theta}.
					\end{equation}
     
					A vague prior distribution $\boldsymbol{\theta} \sim P()$ is then assigned to this parameter.

    The CP model implements the maximum sharing of information, serving as a polar opposite reference to the IP model, enabling an assessment of the assumption of equal effects between indications.

				\paragraph{Mixed common and independent parameters (MCIP) model}
					
				    Here, a mixture of common and independent parameters (MCIP) is assumed.  
                        This is facilitated by a mixture indicator variable, $c_j$, which follows a Bernoulli distribution. In each iteration of the MCMC simulation, this variable is either equal to 1, in which case the $\boldsymbol{\theta}_j$ are assumed equivalent, or is equal to 0, in which case $\boldsymbol{\theta}_j$ are assumed independent, 
					\begin{equation}
						\begin{aligned}
						  &\boldsymbol{\theta}_j = \begin{cases}
						    \boldsymbol{\theta},                  \quad &c_j = 1 \\
						    \boldsymbol{\theta}_j \sim P(), \quad &c_j = 0
						    \end{cases} \\
						  &c_j \sim Bernoulli(p_j).
						  \end{aligned}
					\end{equation}
    					The mixture hyper-parameter, $p_j$, is assigned a vague beta prior distribution, $p_j \sim \beta(1, 1)$ to reflect uncertainty in its values.  
                            The posterior mean of the mixture indicator, $c_j$, represents the mixture probability quantifying the level of mixing. 
                        
                        The MCIP model combines the IP and CP models, providing additional flexibility for the data to determine, for each indication, the plausibility of the IP and CP assumptions. Note that, in this model, the overall pooled parameter, $\boldsymbol{\theta}$, does not represent a common effect across all indications in the data set, but rather a common effect conditional on such a common effect existing.  For example, an indication with data that is extreme in relation to those of other indications will be estimated to have a small mixture probability value and, for this reason, will make only a negligible contribution to the overall pooled parameter estimate.

				\paragraph{Random parameters (RP) model}
					
					Here, the indication-level parameters are assumed to be fully exchangeable (also termed random), and vary according to a common distribution,
					\begin{equation}
						\boldsymbol{\theta}_j|\boldsymbol{\eta}_j \sim F(\boldsymbol{\eta}_j).
					\end{equation}
					The exchangeability distribution is typically defined as normal. For the univariate between-indications model, it is defined as 
					\begin{equation} \label{eqn:univariateExchangeabilityDistribution}
						d_j|m_d, \tau_d \sim F(m_d, \tau_d) = N(m_d, \tau_d^2),
					\end{equation}
					where $m_d$ is the overall pooled effect and $\tau_d$ is the between-indications standard deviation.
					A vague normal prior distribution is placed on the pooled parameter $m_d \sim N(0, 10^2)$, and a weakly-informative half-normal prior distribution is placed on the standard deviation parameter $\tau_d \sim |N(0, 0.5^2)|$.

     For the bivariate between-indications model, the exchangeability distributions are given by,
					\begin{equation} \label{eqn:surrogacyExchangeabilityDistribution}
						\begin{aligned}
							&\lambda_{0j}|\beta_0, \xi_0 \sim  F(\beta_0, \xi_0) = N(\beta_0, \xi_0^2) \\
							&\lambda_{1j}|\beta_1, \xi_1 \sim  F(\beta_1, \xi_1) = N(\beta_1, \xi_1^2) \\
							&\psi_j|h \sim F(h) = |N(0, h)|
						\end{aligned}
					\end{equation}
					where $\beta_0$ and $\beta_1$ are the overall pooled intercept and slope parameters, respectively, $\xi_0$ and $\xi_1$ are the associated between-indications standard deviation parameters, and $h$ is the between-indications variance of the conditional variances.  The overall pooled intercept and slope parameters can be assigned a vague normal prior distribution $\beta_0, \beta_1 \sim N(0, 10^2)$, and a weakly-informative half-normal prior distribution can be placed on the between-indications standard deviation parameters $\xi_0, \xi_1 \sim |N(0, 0.5^2)|$.  Finally, a vague gamma prior distribution can be placed on the between-indications variance parameter for the conditional variances $h \sim \Gamma(1, 0.01)$.

     The RP model assumes full exchangeability, and the data determines the level of sharing via the parameters of the common distribution. The between-indications standard deviation parameters quantify the level of heterogeneity between the indication-level effect estimates. Smaller standard deviation values suggest that the effect estimates are expected to be more similar across indications (with results of the RP model approximating  those of the CP model), and larger values suggest that the effect estimates differ significantly from each other (with results of the RP model approximating  those of the IP model).  Within this model, indication-specific effects are shrunk towards the overall mean effect. The degree of shrinkage depends on  the data. An extreme but imprecise IP model effect is likely to be significantly shrunken in the RP model  towards the overall mean, whilst retaining large uncertainty. A less extreme and imprecise IP model effect may become more precise under the RP model, but with the point estimate not significantly changed.

				\paragraph{Mixed random and independent parameters (MRIP) model}
					
					Here, a mixture of random and independent parameters (MRIP) is assumed 
					\begin{equation}
						\begin{aligned}
						  &\boldsymbol{\theta}_j \sim \begin{cases}
						    F(\boldsymbol{\eta}_j), \quad &c_{j} = 1 \\
						    P(), \quad &c_j = 0
						    \end{cases} \\
						  &c_j \sim Bernoulli(p_j) 
						\end{aligned}
					\end{equation}
					where $c_j$ and $p_j$ are the mixture indicator and mixture hyper-parameter variables, respectively, and have the same interpretation as in the MCIP model. Here, $F(\boldsymbol{\eta}_j)$ represents the exchangeability distribution which is the same as in the RP model (Equations \ref{eqn:univariateExchangeabilityDistribution} and \ref{eqn:surrogacyExchangeabilityDistribution}), and $P()$ is the set of vague prior distributions which are the same as those defined for the IP model.
					 
      The MRIP model provides additional flexibility in relation to the RP model for the data to determine, for each indication, the plausibility of the RP assumption via the mixture indicator. In a similar way to the MCIP model, the parameters describing the exchangeability distribution in the MRIP model do not represent a common distribution across all indications in the data set. Instead, they represent a common distribution across indications where a common distribution is deemed to exist (as determined by the data).
      
     Within this model, $\boldsymbol{\theta}_j$ represents the indication-specific estimate which is a weighted-average (i.e., weighted by the mixture probability) of the exchangeable and independent components.
     For the univariate model, $d_j$ is the indication-specific treatment effect estimate which is informed by sharing information across indications (via the exchangeability distribution), where the degree of sharing is regulated by the mixture probability.
		
			\subsubsection{Obtaining estimates of a treatment effect on overall survival (OS)} \label{sec:prediction}
			
				In this section, we will consider two points which must be addressed when using the models described above to predict treatment effects in the context of HTA. Firstly, how to obtain indication-specific parameter estimates from each of the sharing models (from the between-indications component). Secondly, how to obtain predictions of a treatment effect on OS from the bivariate models.

    Considering the between-indications component, the indication-level parameters in the IP model draw only from the indication-specific evidence. Because sharing is not allowed, this model does not produce a summary estimate across all indications, or an estimate for a new indication. 
    All remaining models impose sharing across indications and so the indication-specific estimates are  influenced by both direct (within-indication) and indirect (between-indications) evidence.  In the CP model, the overall pooled estimate reflects the expected value (for the effect on a single endpoint or on a surrogacy parameter) in all of the observed indications (which are assumed equal). 
    For the univariate models, this overall pooled estimate represents both the treatment effect across all indications and the effect for a new indication (since these are assumed to be equivalent).
                
                From the RP model, appropriate estimates for observed indications are the shrunken estimates $\boldsymbol{\theta}_j$. For a new indication, an estimate can be predicted by sampling from the predictive distribution. For example, for the univariate RP and MRIP models, the predictive distribution would be given by $d_{pred} \sim N(m_d, \tau_d^2)$. In the HTA context, policy decisions are made for a population and the estimates used to inform decision-making must take heterogeneity in the treatment effects into consideration. As such, the mean of the common distribution in the RP model could lead to overprecise estimates of treatment effect and the predictive distribution is recommended as a basis for decision-making \cite{welton2007correction}.
                
				For the MCIP and MRIP models, indication-specific estimates are based on the mixture probability, the  estimate from the sharing component  (respectively, CP and RP) and the independent component estimate. For a new indication, a mixture probability needs to be specified separately to allow for predictions from the sharing component and from an independent component (which, in the absence of evidence, could be drawn from a vague prior distribution). Where the new indication is considered perfectly exchangeable with the indications deemed similar by the model (i.e. mixture probability equal to one) predictions from only the sharing component of these models can be used.
				
					Different approaches can be used to obtain predictions of a treatment effect on OS in a target indication using the bivariate models, i.e. utilising indication-level parameter estimates obtained from the between-indications component of the above models for sharing of information on the intercept ($\lambda_{0j}$), slope ($\lambda_{1j}$), and conditional variance ($\psi_j^2$)  parameters (see Equation \eqref{eqn:surrogacyBetweenStudies} in Section \ref{sec:likelihood}).

                   Note that, under the formulation by Daniels and Hughes, the bivariate models assume independent parameters  for the surrogate endpoint (PFS) effect at the study-level. 
Consequently, the surrogate relationship estimated by the bivariate model  needs to be applied to an indication-specific effect estimate for PFS to predict an effect on OS. Decisions over the sharing of information on this indication-specific PFS effect estimate can differ from those on the bivariate model into which it is entered. We implemented two options. The first enters the indication-level PFS estimate from the univariate IP model into the indication-specific surrogate relationships estimated by each of the bivariate sharing models (CP, MCIP, RP, MRIP). In this case, the predicted effect on OS will be based on sharing information on the  surrogate relationships but not on the effects on PFS across indications. The second approach involves matching the between-indications relationship to share on both effects on PFS and surrogacy parameters (i.e., the PFS effect estimate from the CP univariate model is entered into the surrogate relationship estimated by the CP bivariate model, etc.). 
Formulae for obtaining a predicted treatment effect on OS from an effect estimate on PFS are described in Appendix \ref{sec:predictionFormulae}.

		\subsection{Model implementation} \label{sec:modelImplementation}
			
			\subsubsection{Application to case study}
			The proposed methods will be initially applied to data on bevacizumab available at present day (see Figure \ref{fig:observedPresentDay}), across all indications. Given that bevacizumab was one of the earliest multi-indication oncology drugs, this analysis will examine a well developed evidence base (as well developed as can be expected in HTA).  Results from the present day analysis will represent a benchmark with which to compare the results from the more policy-relevant analyses at (earlier) appraisal time points, where less evidence had accumulated across indications. The analyses at two particular appraisal time points (TA178 in RCC,  and TA285 in OFTPP cancer) assume we want to estimate effects for the target indication in those appraisals, using the multi-indication data available at the time of each appraisal (see figures in Appendix \ref{sec:appraisalDataSets} for a summary of each data set). These  analyses will illustrate whether efficient estimates of key parameters of interest for HTA decision-making can still be obtained  with less developed evidence bases, by sharing information across indications.

			\subsubsection{Software}
				All models were implemented in OpenBUGS, via the \textit{R2OpenBUGS} package (version 3.2.1) \cite{Sturtz2019} in R (version 4.1.3) \cite{R2021}, using MCMC sampling to estimate Bayesian posterior distributions for model parameters.
				Each implementation consisted of three MCMC chains, where 20,000 iterations were used as an initial burn-in period for each chain.
				Posterior estimates were based on 80,000 samples per chain, and checked for sensitivity to changes in initial values.
				The effective sample size and $\hat{R}$ statistics were used to assess non-convergence of chains \cite{Gelman1992}.	
    
			\subsubsection{Assessment of fit and model choice}
	The deviance information criterion (DIC) provides a measure of how well a model fits a data set, whilst penalising model complexity in terms of the number of parameters, where lower values are preferred. This measure, along with the residual deviance, are useful for comparing different models when applied to the same data set \cite{Spiegelhalter2002}.
				In our application, we use the residual deviance and DIC to select between the multi-indication meta-analysis models (IP, CP, MCIP, RP, MRIP), for the univariate and bivariate models separately, at each analysis time point.
			We use a difference of three units in DIC to represent a meaningful difference in model fit between two models, as recommended by Spiegelhalter \textit{et al} \cite{Spiegelhalter2002}. Where the difference is less than three units in DIC, the simplest model in terms of the number of effective parameters is selected.
		
	\section{Results} \label{sec:results}
		
		We report the results from applying the multi-indication meta-analysis methods described in Section \ref{sec:methods}, for sharing of information either on treatment effects on an individual endpoint or on a surrogate relationship between effects on two endpoints, to the data available at the following three time points (see Figure \ref{fig:taTimelineSummary} for an illustration of trial data with respect to these time points); present day,  TA178 (RCC), and TA285 (OFTPP cancer).

		\subsection{Present day analyses}

\begin{figure}[!htb]
				\centering
				\includegraphics[width=\textwidth]{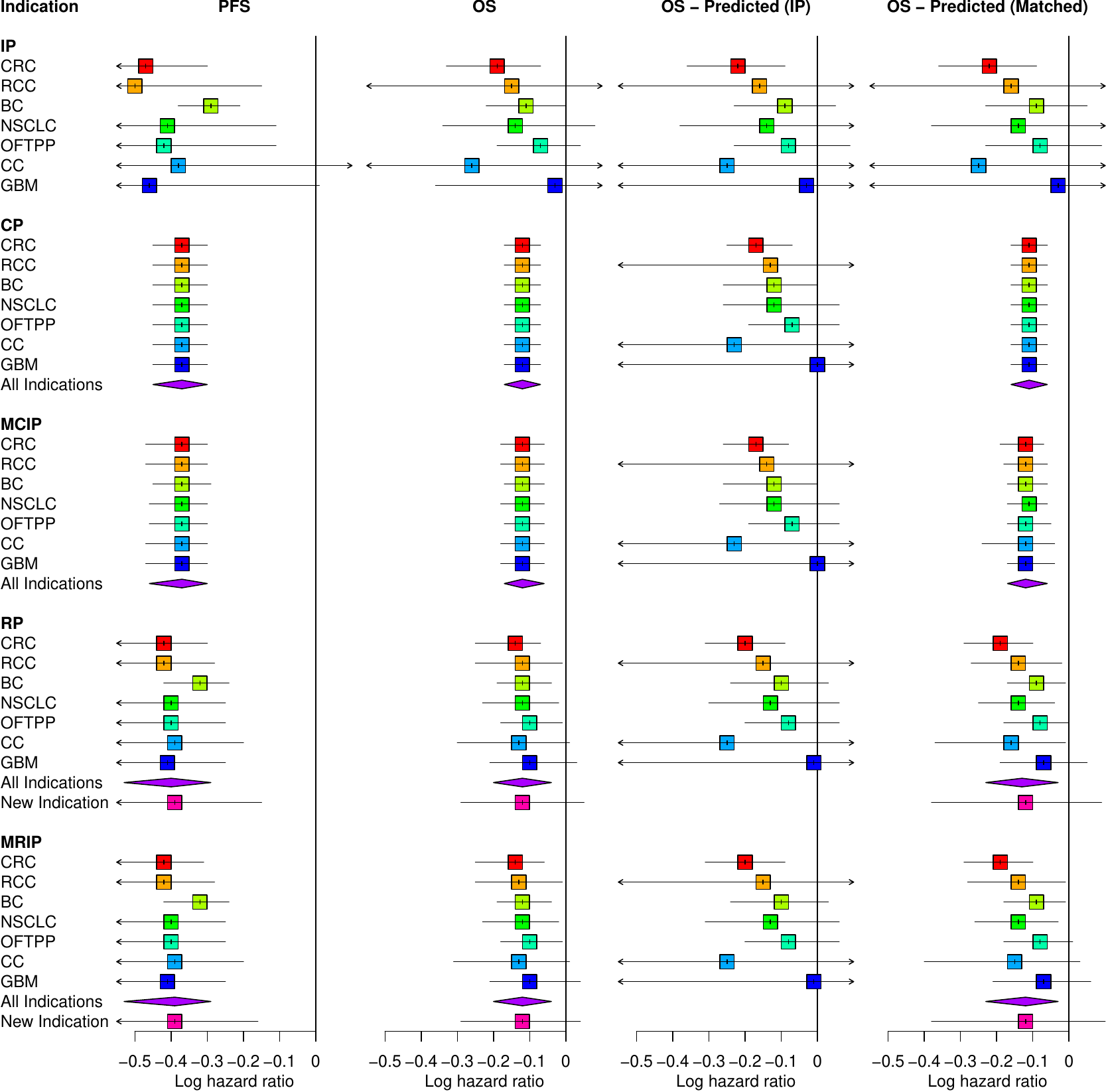}
				\caption{Treatment effect estimates (on the log hazard ratio scale) from multi-indication meta-analysis of data available at present day. For the CP model, the indication-level estimates are all equivalent to the 'All indications' estimate, on the PFS, OS, and OS - Predicted (Matched) outcomes. PFS - progression-free survival, OS - overall survival, IP - independent parameters, CP - common parameter, MCIP - mixed common and independent parameters, RP - random parameters, MRIP - mixed random and independent parameters, CRC - colorectal cancer, BC - breast cancer, NSCLC - non-small cell lung cancer, RCC - renal cell carcinoma, CC - cervical cancer, GBM - glioblastoma, OFTPP - ovarian, fallopian tube, and primary peritoneal.}
				\label{fig:presentDayAllResults}
			\end{figure}

			Figure \ref{fig:presentDayAllResults} illustrates the results of the analyses of data available at present day. It presents the median and 95\% credible interval (CrI) for the treatment effect estimates corresponding to each indication, grouped by the between-indications model. The columns denoted as `PFS' and `OS' include the estimates from the application of the univariate models to synthesise effects on PFS and OS, respectively. The predicted effects on OS obtained from the application of the bivariate models are included in the columns identified as `OS - Predicted (IP)' and `OS - Predicted (Matched)'. We will describe the results from the univariate and bivariate models in turn.

    \subsubsection{Results from univariate analyses of treatment effects on individual endpoints}

In line with the trial-level results (presented in Figure \ref{fig:observedPresentDay}), univariate synthesis of treatment effects on PFS suggests bevacizumab to be beneficial across the majority of indications, with no evidence of heterogeneity in the effects across indications. Where within-indication evidence (IP model) is insufficient for the treatment effect estimates to be statistically meaningful (e.g. for CC and GBM), sharing of information across all indications increases the precision of the estimates leading to CrIs that do not include the null effect (CP, MCIP, RP, MRIP models). 
                
                A  positive treatment effect on OS is also apparent from within-indication evidence (IP model) for those indications with more data, e.g., CRC (nine trials) and BC (ten trials), but data are too limited to make conclusions for the other indications. Thus, there is  no suggestion from these data that sharing on OS is inappropriate. 
            
              Estimates from the CP model, which assumes equality of the treatment effects and provides maximal sharing across indications, suggest that a meaningful positive effect can be expected on both outcomes.
                The RP model, which allows for variability in effects across indications, provides indication-specific estimates that are more similar to each other compared to those from the IP model (i.e. they are shrunken towards the overall mean), but retain wider uncertainty for the indications with sparser data (e.g., CC or GBM). Notably, the point estimate of the PFS effect for BC has not been substantially shrunken (due to the higher level of evidence for this indication and  discrepancy with the point estimates from the other indications).
               
                The estimates obtained from the mixture models are in close agreement with those from the corresponding non-mixture sharing model (i.e., MCIP with CP, MRIP with RP), and the mixture probabilities are approximately equal to one for most indications (see Table \ref{tab:mixProbsPresentWithBranch} for the mixture probability estimates). These results add plausibility for sharing across indications in this case study as the data are relatively consistent. The  discrepancy in the point estimate for the PFS effect in BC is accounted for by the MCIP model, by attributing this indication a  lower mixture probability (mean 0.96, standard deviation 0.19) compared to the other indications (see Table \ref{tab:mixProbsPresentWithBranch}).
                There is no meaningful difference in the goodness-of-fit between the different models for the synthesis of treatment effects on both OS and PFS, as the DIC values (presented in Appendix \ref{sec:appPresentDic}) differ by less than 3 units.
                This suggests that whilst any model could be considered appropriate, the application of the simpler CP model is most efficient in describing these data.

\begin{table}[H]
\centering
\caption{Mixture probability estimates (as mean and standard deviation), and informative and vague mixture component estimates (as median and 95\% credible interval), from the application of univariate models to data on overall survival (OS), and progression-free survival (PFS), at present-day. } 
\label{tab:mixProbsPresentWithBranch}
\begin{tabular}{llrrr}
  \toprule
Model & Indication & Probability & Informative & Vague \\ 
  \midrule
\textbf{OS} &  &  &  &  \\ 
  MCIP & BC & 1.00 (0.05) & -0.12 (-0.17, -0.06) & -0.10 (-61.80, 61.70) \\ 
  MCIP & CC & 0.99 (0.11) & -0.12 (-0.17, -0.06) & -0.21 (-61.70, 61.85) \\ 
  MCIP & CRC & 0.99 (0.08) & -0.12 (-0.17, -0.06) & -0.14 (-62.23, 61.94) \\ 
  MCIP & GBM & 0.99 (0.08) & -0.12 (-0.17, -0.06) &  0.02 (-62.00, 61.71) \\ 
  MCIP & NSCLC & 1.00 (0.06) & -0.12 (-0.17, -0.06) & -0.13 (-61.95, 62.01) \\ 
  MCIP & OFTPP & 1.00 (0.07) & -0.12 (-0.17, -0.06) & -0.04 (-61.79, 61.94) \\
\bottomrule
\end{tabular}
\end{table}
                \clearpage

\subsubsection{Results of surrogate endpoint evaluation }
\begin{figure}[!htb]
				\centering
				\includegraphics[width=\textwidth]{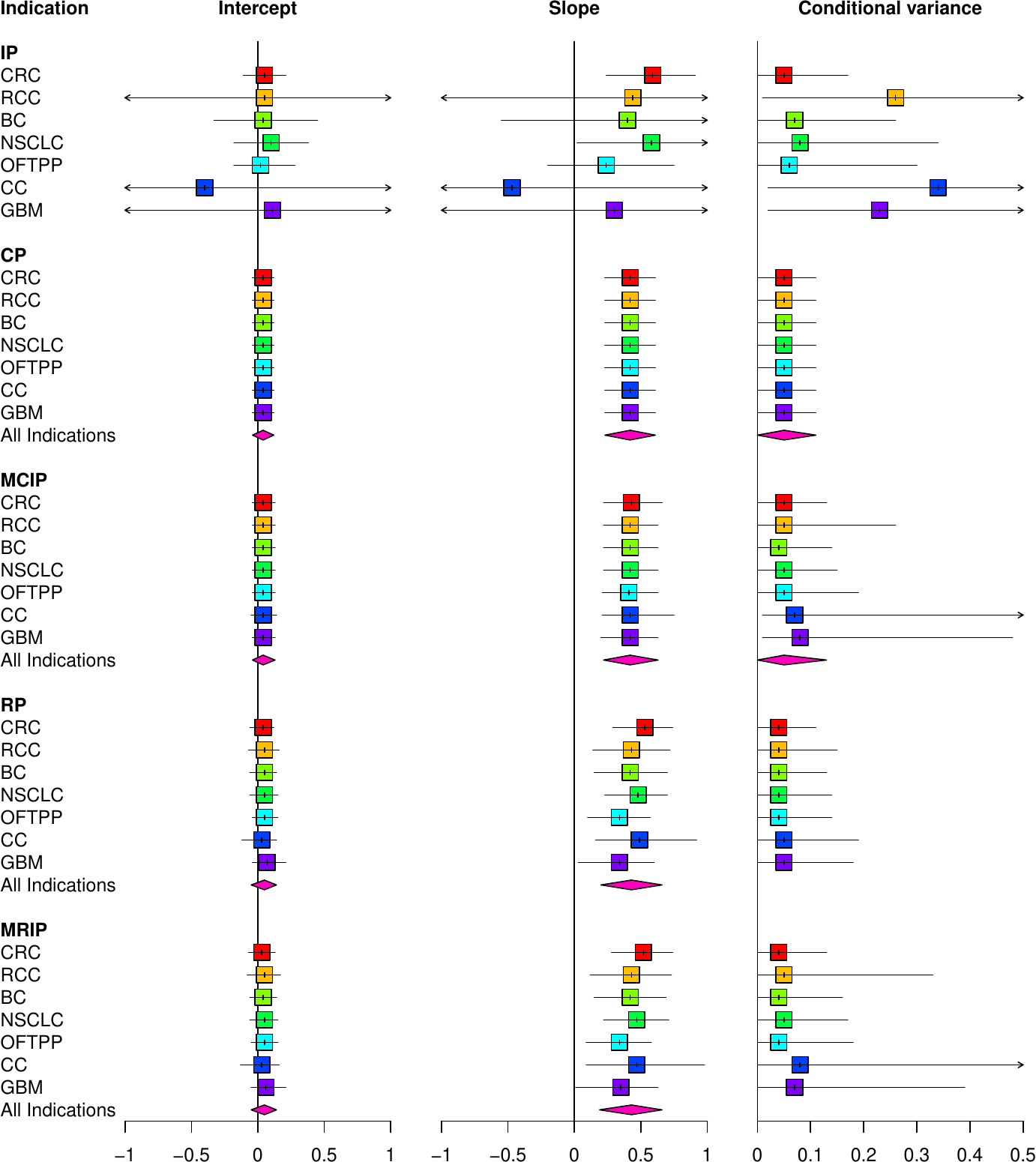}
				\caption{Multi-indication meta-analysis of surrogacy parameters using the data available at present day. For the CP model, the indication-level estimates are all equivalent to the 'All indications' estimate. IP - independent parameters, CP - common parameter, MCIP - mixed common and independent parameters, RP - random parameters, MRIP - mixed random and independent parameters, CRC - colorectal cancer, BC - breast cancer, NSCLC - non-small cell lung cancer, RCC - renal cell carcinoma, CC - cervical cancer, GBM - glioblastoma, OFTPP - ovarian, fallopian tube, and primary peritoneal.}
				\label{fig:presentDaySurrogacy}
			\end{figure}

Before interpreting the results from using the bivariate models to predict a treatment effect on OS in Section \ref{sec:predictedEffectsOnOs}, we focus on the inferences obtained for the parameters describing the surrogate relationship (i.e., intercept, slope, and conditional variance parameters) between treatments effects on PFS and OS. Figure \ref{fig:presentDaySurrogacy} depicts the median and 95\% CrI estimates for these parameters based on the present day data set, where each column corresponds to a particular parameter and each row includes the indication-specific estimate. 
The estimates are grouped by the between-indications model.
Further results of this analysis can be found in Appendix \ref{sec:appPresentSurrogacy}, and Appendix  \ref{sec:appPresentCrossVal} presents the results from a cross validation procedure used to evaluate the surrogate relationship in more detail \cite{Bujkiewicz2019a}. Note that the intercept parameter is estimated on the same  scale as a treatment effect on the final outcome (OS), and can be interpreted as a  log HR.
The slope parameter quantifies how the effect on the final outcome, OS, varies with respect to the effect on the surrogate endpoint, PFS, and as such can be interpreted as a ratio of log HRs.
                
                The results from applying the IP model show that the surrogacy parameters are estimated with  large uncertainty in indications with limited data; RCC (two trials), CC (one trial), and GBM (three trials). Consequently, it is difficult to determine whether the surrogacy criteria (described in Section \ref{sec:likelihood}) are satisfied for these indications by considering the within-indication evidence alone. Across all indications, CrIs for intercepts contain the value zero (as desired according to the surrogacy criteria).  The CrIs for the slope parameter for CRC and NSCLC (indications with data on both PFS and OS effects from 6,067 patients across nine trials, and 3,034 patients across six trials, respectively) do not include zero.  This suggests a positive association between the treatment effects on the two outcomes within these indications,  where a larger effect on PFS is associated with a larger effect in OS, and is indicative of a strong surrogate relationship.  
                However, the CrIs corresponding to the slope parameter for BC and OFTPP cancer include zero, despite the availability of a similar level of data (from 4,459 patients across ten trials in BC, and 4,345 patients across seven trials in OFTPP cancer).  This absence of evidence does not imply that surrogacy does not hold within these indications, but encourages caution in the  interpretation of the results from the less flexible sharing models, particularly those assuming common or exchangeable surrogacy parameters across all indications (CP and RP).

                 The conditional variance estimates were relatively large for RCC, CC and GBM (compared to other indications), reflecting the sparse evidence for determining  surrogacy  in these indications.  Out of the remaining indications (BC, NSCLC, OFTPP cancer and CRC), CRC presents the smallest conditional variance estimate with the narrowest CrI, due to the higher level of evidence for the existence of a surrogate relationship (with precise estimates for all three surrogacy parameters).  The higher uncertainty in NSCLC and OFTPP cancer estimates (in relation to CRC) may be due to the relatively small number of studies; six and seven studies, respectively. In contrast, the higher uncertainty in BC estimates (in relation to CRC), where the number of studies was ten, is indicative of a weaker surrogate relationship within this indication. The point estimates of the conditional variance parameter corresponding to BC, NSCLC and OFTPP cancer all appear to be larger compared to the point estimate for CRC, but this may be due to the higher uncertainty in these indications and a positively-skewed distribution for this parameter.  
                
                Indication-specific estimates from the sharing models are more precise compared to those from the IP model. For instance, the slope parameter estimates (including CrIs) are positive in across indications for all sharing models.  An exception to this are the conditional variance estimates from applying the mixture models (MCIP and MRIP), which still show relatively large uncertainty for the indications with sparser data: RCC, CC, and GBM. Mixture probabilities for the intercept and slope parameters are all approximately equal to one  (see Appendix \ref{sec:appPresentSurrogateMixProbs}),  but range between 0.57 and 0.82 for the conditional variance parameter, assuming lowest values in CC and GBM. 

                In terms of goodness-of-fit (see Appendix \ref{sec:appPresentSurrogateDic}), the mixture models have notably smaller DIC values compared to the non-mixture models, with the lowest DIC value for the MRIP model.  In this application, the added flexibility of the mixture models (i.e. the ability to regulate the degree of sharing for each indication) is likely to be more suitable for  sharing on the conditional variance parameter as the corresponding mixture probabilities are lower and more variable. This indicates a lack of support from the data for the assumptions of equivalence or exchangeability across indications on this parameter. 

\subsubsection{Predicted effects on overall survival (OS) using surrogate relationships } \label{sec:predictedEffectsOnOs}
               The predicted treatment effects on OS obtained from the bivariate models for each indication are depicted in  Figure \ref{fig:presentDayAllResults}, in columns `OS - Predicted (IP)' and `OS - Predicted (Matched)' (see Section \ref{sec:prediction} for a description of how these are obtained).
                
                The predicted OS effects estimated by the bivariate IP model are consistent with the pooled OS effects estimated by the univariate IP model, albeit with slightly larger uncertainty (see Table \ref{tab:predOsIpPresent}  in Appendix \ref{sec:appPresentPredicted} for specific numerical values). The predicted OS effects estimated by the bivariate models which allow for sharing of information on the surrogacy parameters (but without sharing on the PFS effect) are only slightly more precise compared to estimates from the bivariate IP model, even when implementing a CP assumption. This demonstrates the limited impact of  sharing on the surrogate relationship alone (when not sharing on the PFS effect).  It is only when sharing on both the PFS effect and the surrogate relationship  (OS - Predicted (Matched) column) that notable precision gains are obtained in relation to not sharing.  This suggests that sharing on the PFS effect, to obtain a more precise pooled estimate, is particularly important in predicting the corresponding OS effect using the bivariate models.
                
                Considering the matched bivariate models, which allow for sharing on both the PFS effect and the surrogate relationship (by matching the sharing assumption made on the two sets of parameters), the predicted estimates from the  bivariate CP model show the lowest uncertainty, which is to be expected as this model allows maximal sharing. These estimates are consistent with the pooled OS estimates from the univariate CP model.   The predicted estimates from the bivariate MCIP model show wider uncertainty than CP model estimates. The bivariate RP and MRIP models present more variation in the point estimates across all indications, all being in line with the IP model but shrunken towards the overall mean value.  The shrinkage is greater for indications other than CRC and BC.  

  The level of uncertainty in the OS effect estimates predicted by the matched bivariate sharing models is similar to the uncertainty in the pooled OS effects estimated by the univariate sharing models.
  However, there are some differences in the estimates between the univariate and bivariate approaches. For instance, the bivariate approach leads to higher variation in point estimates across indications. Furthermore, predictions for a new indication made by the hierarchical sharing models (RP and MRIP) have larger uncertainty in the bivariate context.  For CC, for which data were only available from a single trial, the RP model CrIs include the null effect when sharing across indications on OS alone, but  not when predicting the OS effect by sharing on both the PFS effect and the surrogacy parameters (see CrI in OS - Predicted (Matched) column).

				\subsection{Appraisal specific analyses}
\begin{figure}[!htb]
				\centering
				\includegraphics[width=\textwidth]{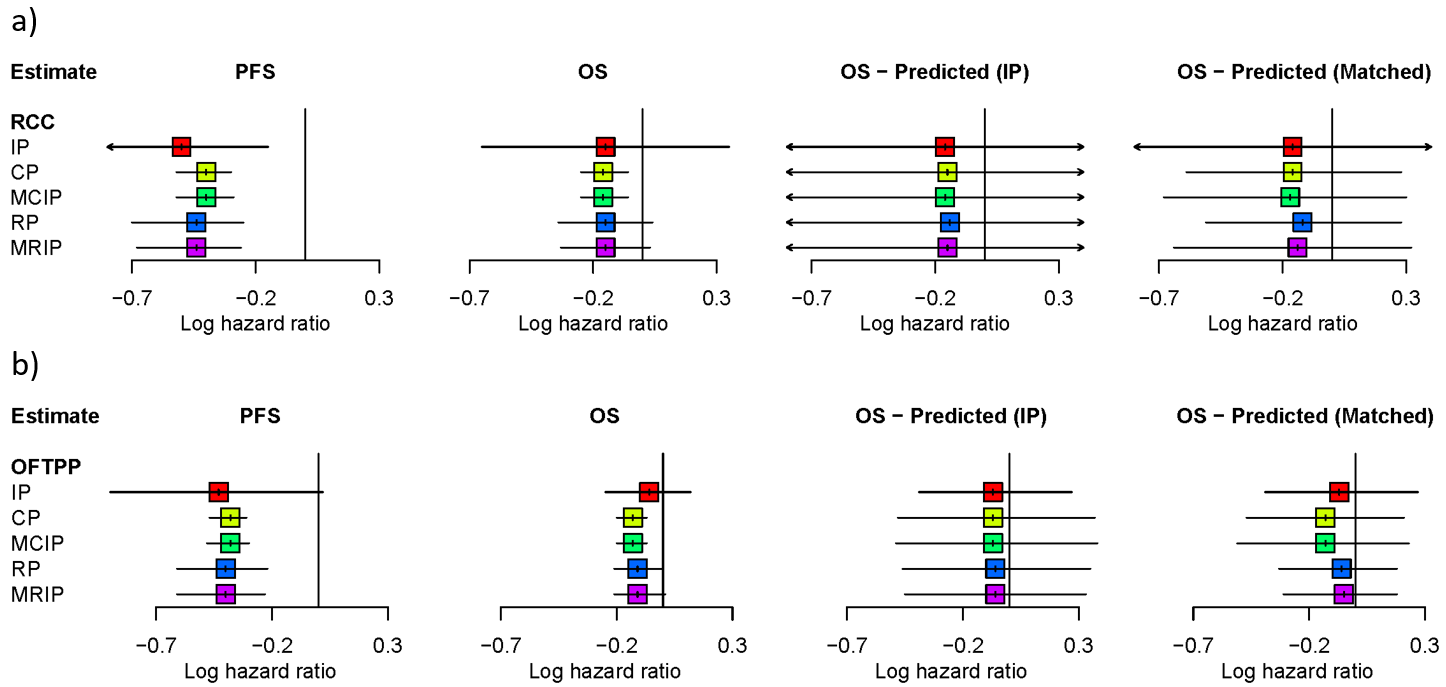}
				\caption{Multi-indication meta-analysis of data available at the time of: a) TA178 (target indication: renal cell carcinoma (RCC)), and b) TA285 (target indication: ovarian, fallopian tube, and primary peritoneal (OFTPP) cancer). PFS - progression-free survival, OS - overall survival, IP - independent parameters, CP - common parameter, MCIP - mixed common and independent parameters, RP - random parameters, MRIP - mixed random and independent parameters, RCC - renal cell carcinoma, OFTPP - ovarian, fallopian tube, and primary peritoneal.}
				\label{fig:taAllResults}
			\end{figure}

			Figure \ref{fig:taAllResults} depicts the results obtained from perfoming  multi-indication meta-analyses using the data that were available on PFS and OS at the time of two NICE TAs; TA178 for RCC (panel `a' in Figure \ref{fig:taAllResults}), and TA285 for OFTPP cancer (panel `b'). Further results, including numerical estimates, are presented in Appendix  \ref{sec:appTa178Pooled} and Appendix \ref{sec:appTa285Pooled}) respectively.
			Broadly, the point estimates obtained from the different models are consistent with one another within the target indication, and there is large overlap in the CrIs for the effects on both PFS and OS.
		The estimates from the sharing models (CP, MCIP, RP, MRIP) show improved precision compared to the IP model, which does not allow for sharing across indications.
            
            The results from the TA time point analyses show a similar trend to those obtained from the analysis of present day  data, in terms of the benefit of sharing across indications for gaining precision in effect estimates.
            However, less evidence had accumulated at these earlier time points and so the corresponding analyses provide less precise predicted OS effect estimates (using both the univariate and bivariate approaches).
            In particular, the predicted OS effect estimates obtained from the bivariate models are noticeably more imprecise compared to the pooled estimates from the univariate models.
            This remains the case even when sharing is imposed on both the PFS effect and the surrogate relationship, which contrasts with the results from the present day analysis, suggesting that predictions based on the bivariate models are unlikely to strengthen estimates in realistic appraisal contexts.

	\section{Discussion} \label{sec:discussion}
		
		In this paper, we explored alternative meta-analysis models for the synthesis of evidence on treatment effects on an individual endpoint, or on a surrogate relationship between treatment effects on a surrogate endpoint and a final clinical outcome, across multiple indications in oncology.
		We demonstrated the methods using data on treatment effects of bevacizumab on PFS and OS from trials across multiple cancer indications, by applying them to all data available at present day and, in the context of HTA, to data that had accumulated at time points corresponding to two NICE TAs of bevacizumab; for renal and ovarian cancers.
                
Bevacizumab is one of the first examples of a targeted multi-indication cancer drug, and therefore it's evidence base is relatively well developed at present day. Such evidence still shows considerable uncertainty in the OS effect estimates within the majority of the indications for which bevacizumab is licensed, but does not show any noticeable heterogeneity in effects across indications (see Figure \ref{fig:observedPresentDay}).   This suggests that a case study on bevacizumab is appropriate to explore whether sharing information across indications has the potential to strengthen indication-specific evidence, for inference in the HTA context.

Our case study showed that the application of methods for sharing information across indications can lead to more precise inferences on treatment effects, and on the parameters describing a surrogate relationship. This gain in precision was more noticeable for indications with limited evidence. Although the OS effects predicted using the surrogacy parameters were consistent with the pooled OS effects (from sharing across indications on OS effects alone), they only showed meaningful gains in precision when sharing was imposed on the PFS effects in addition to the surrogacy parameters.  Corresponding analyses using the earlier, sparser (within and across indications) evidence available at particular HTA appraisal points showed that sharing on surrogacy parameters, even when also sharing on PFS effects, did not notably reduce uncertainty in predicted OS effects. These findings highlight that uncertainty in the surrogacy parameter estimates may be due to limited evidence or the lack of a clear surrogate relationship within the broader evidence base (i.e., within other indications), and this must be considered when determining whether sharing across indications on these parameters is appropriate.

There were differences in the support provided by the indication-level evidence for surrogacy between indications with higher levels of evidence.
For example, the evidence suggested a surrogate relationship within CRC (nine trials); with small, precise estimates for the intercept and conditional variance parameters, and a precise non-zero estimate for the slope parameter.
Such estimates are indicative of surrogacy, although there are no clear cut rules to infer this statistically.
Despite there being a similar level of evidence in BC (ten trials), there was higher uncertainty in the slope and conditional variance parameters which implies that surrogacy may not hold within this indication.
Thus, the surrogacy parameter estimates based on sharing information from BC may be deemed overly precise, even for the mixture models which provide greater flexibility to regulate the level of sharing from a particular indication.
                However, it can be argued that as long as these results are not used to judge the strength of the surrogate relationship within BC alone, this sharing can be implemented to improve our predictions of the effect on the final outcome in particular indications.
                We performed a sensitivity analysis by applying the sharing models to the present day data after removing trials in BC but did not see a notable difference in the surrogacy parameter estimates in the remaining indications (see Appendix \ref{sec:appPresentSurrogacyBreast}).

The predicted OS effect estimate was more greatly influenced by the precision of the corresponding PFS effect estimate, which was entered into the bivariate model to make the prediction, than the surrogacy parameter estimates themselves.
This suggests that applying a bivariate model to predict an OS effect may only be applicable where there is strong evidence on the PFS effect.
In a HTA context, the feasibility of predicting the OS effect for a particular indication will depend on the level of evidence available on PFS within that indication and across other indications from which sharing is deemed plausible.
Further research is required to explore the conditions under which  sharing on surrogate relationships strengthens predictions of OS effects. We expect that this is likely to require that at least one indication in the broader, multi-indication, evidence base has a meaningful level of evidence (in terms of data on treatment effects on both the surrogate endpoint and the final outcome) and shows support for surrogacy.

In this work, we explored a range of models imposing different levels of information sharing, including no sharing across indications, models with increasing levels of flexibility in the strength of sharing, to assuming an equal, common, effect across all indications. These models fit the data similarly when sharing on each endpoint individually (univariate models), adding to the plausibility of sharing across all indications, and suggesting that the added flexibility of the models with higher parametrisations was not required.
However, the application of the more flexible sharing (mixture) models requires important methodological considerations when sharing on surrogate relationships (bivariate models). 
For instance, the mixture probabilities were much lower for the conditional variance parameter, compared to the intercept and slope parameters, cautioning against sharing on this parameter.
This may also explain why the mixture models provided a better fit to these data compared to the non-mixture models which make more rigid assumptions regarding sharing.
We performed a sensitivity analysis where the sharing models were implemented with the assumption of independent conditional variance parameters (i.e., sharing on the intercept and slope parameters only), but did not see a notable difference in results (see Appendix \ref{sec:appPresentSurrogacyIndependentPsi}).
A better understanding of the scenarios where the mixture models offer a benefit over the non-mixture models, for example, where data from a particular indication are more extreme relative to the other indications, would be provided by a simulation study. Some of such scenarios for sharing information on surrogate relationships across treatment classes have been explored by Papanikos \textit{et al} \cite{Papanikos2020}.
A simulation study could also provide a systematic assessment of the models' performance (i.e., when they are prone to bias, or over/under coverage, with respect to a known true effect).
                
 To illustrate and explore the sharing methods in a multi-indication evidence synthesis context, in our case study, we considered it appropriate to focus the synthesis on aggregate trial-level data on a treatment effect for a survival outcome, represented by a log HR on either PFS or OS.
                However, future research on multi-indication meta-analyses should extend the methodological framework to account for scenarios where the log HR is not representative of the true effect; for example, where there is evidence of non-proportional hazards between trial arms.
		 
Meta-analytic methods can be usefully applied to share information across indications to obtain more precise treatment effect estimates on an endpoint of interest.
Such an approach is also applicable when using a bivariate model to predict an effect on a final outcome based on the effect on a surrogate endpoint, where sharing across indications can provide precision gains for both the effect on the surrogate endpoint and the parameters quantifying the surrogate relationship.
Mixture models may be more appropriate when sharing on a surrogate relationship, where greater flexibility is required to regulate the degree of sharing on the conditional variance parameter.
The findings from this case study on bevacizumab are likely to be generalisable to other multi-indication health technologies.
For decision-making in a HTA context, these methods can be beneficial where there is limited evidence on the final outcome within the target indication, and sufficient evidence has accumulated across the other indications to demonstrate the existence of a surrogate relationship.
However, the plausibility of any sharing assumptions should be carefully validated with clinical experts, where possible.

	\section{Acknowledgements}
		This research was funded by the Medical Research Council, Better Methods Better Research panel [MRC MR/W021102/1]. SB and JS were also supported by Leicester NIHR Biomedical Research Centre (BRC). The views expressed are those of the author(s) and not necessarily those of the NIHR or the Department of Health and Social Care.
 
	\bibliographystyle{vancouver}
	\bibliography{singh2023MultiIndication}
	\clearpage

	\appendix
	\section{Additional results} \label{sec:additionalResults}
	\setcounter{table}{0}
	\renewcommand{\thetable}{A\arabic{table}}
	\setcounter{figure}{0}
	\renewcommand{\thefigure}{A\arabic{figure}}
 
		\begin{landscape}
		\subsection{Present day}
			\subsubsection{Pooled OS and PFS estimates}\label{sec:appPresentPooled}
\begin{table}[H]
\centering
\caption{Mean and 95\% credible interval estimates from applying univariate multi-indication meta-analysis models to data available at present-day on treatment effects on overall 
                       survival (log hazard ratio scale).} 
\label{tab:osPresent}
\begin{tabular}{lrrrrr}
  \toprule
Indication & IP & CP & MCIP & RP & MRIP \\ 
  \midrule
CRC & -0.19 (-0.33, -0.07) & -0.12 (-0.17, -0.07) & -0.12 (-0.18, -0.06) & -0.15 (-0.25, -0.07) & -0.15 (-0.25, -0.06) \\ 
  RCC & -0.15 (-0.65,  0.35) & -0.12 (-0.17, -0.07) & -0.12 (-0.18, -0.06) & -0.13 (-0.25, -0.01) & -0.13 (-0.25, -0.01) \\ 
  BC & -0.11 (-0.22,  0.00) & -0.12 (-0.17, -0.07) & -0.12 (-0.17, -0.06) & -0.12 (-0.19, -0.04) & -0.12 (-0.19, -0.04) \\ 
  NSCLC & -0.14 (-0.34,  0.08) & -0.12 (-0.17, -0.07) & -0.12 (-0.18, -0.06) & -0.13 (-0.23, -0.02) & -0.13 (-0.23, -0.02) \\ 
  OFTPP & -0.07 (-0.19,  0.04) & -0.12 (-0.17, -0.07) & -0.12 (-0.17, -0.06) & -0.10 (-0.18, -0.01) & -0.10 (-0.18, -0.01) \\ 
  CC & -0.26 (-1.35,  0.83) & -0.12 (-0.17, -0.07) & -0.12 (-0.18, -0.06) & -0.13 (-0.30,  0.01) & -0.13 (-0.31,  0.01) \\ 
  GBM & -0.03 (-0.36,  0.32) & -0.12 (-0.17, -0.07) & -0.12 (-0.18, -0.06) & -0.10 (-0.21,  0.03) & -0.10 (-0.21,  0.04) \\ 
  All Indications &  & -0.12 (-0.17, -0.07) & -0.12 (-0.17, -0.06) & -0.12 (-0.20, -0.04) & -0.12 (-0.20, -0.04) \\ 
  New Indication &  &  &  & -0.12 (-0.29,  0.05) & -0.12 (-0.29,  0.04) \\ 
   \bottomrule
\end{tabular}
\end{table}
\begin{table}[H]
\centering
\caption{Mean and 95\% credible interval estimates from applying univariate multi-indication meta-analysis models to data available at present-day on treatment effects on progression-free 
                       survival (log hazard ratio scale).} 
\label{tab:pfsPresent}
\begin{tabular}{lrrrrr}
  \toprule
Indication & IP & CP & MCIP & RP & MRIP \\ 
  \midrule
CRC & -0.47 (-0.67, -0.30) & -0.37 (-0.45, -0.30) & -0.38 (-0.47, -0.30) & -0.43 (-0.58, -0.30) & -0.43 (-0.58, -0.31) \\ 
  RCC & -0.53 (-1.01, -0.15) & -0.37 (-0.45, -0.30) & -0.38 (-0.47, -0.30) & -0.43 (-0.61, -0.28) & -0.43 (-0.62, -0.28) \\ 
  BC & -0.29 (-0.38, -0.21) & -0.37 (-0.45, -0.30) & -0.37 (-0.45, -0.29) & -0.32 (-0.42, -0.24) & -0.32 (-0.42, -0.24) \\ 
  NSCLC & -0.42 (-0.73, -0.11) & -0.37 (-0.45, -0.30) & -0.37 (-0.46, -0.30) & -0.40 (-0.58, -0.25) & -0.40 (-0.57, -0.25) \\ 
  OFTPP & -0.41 (-0.71, -0.11) & -0.37 (-0.45, -0.30) & -0.37 (-0.46, -0.30) & -0.40 (-0.57, -0.25) & -0.40 (-0.57, -0.25) \\ 
  CC & -0.38 (-1.48,  0.70) & -0.37 (-0.45, -0.30) & -0.37 (-0.47, -0.30) & -0.39 (-0.61, -0.20) & -0.39 (-0.61, -0.20) \\ 
  GBM & -0.46 (-0.94,  0.01) & -0.37 (-0.45, -0.30) & -0.38 (-0.47, -0.30) & -0.41 (-0.60, -0.25) & -0.41 (-0.60, -0.25) \\ 
  All Indications &  & -0.37 (-0.45, -0.30) & -0.37 (-0.46, -0.30) & -0.40 (-0.53, -0.29) & -0.40 (-0.53, -0.29) \\ 
  New Indication &  &  &  & -0.40 (-0.67, -0.15) & -0.40 (-0.66, -0.16) \\ 
   \bottomrule
\end{tabular}
\end{table}
			\subsubsection{Predicted OS estimates}\label{sec:appPresentPredicted}
\begin{table}[H]
\centering
\caption{Mean and 95\% credible interval estimates for 
                               predicted effects on overall survival (log hazard ratio scale), from entering the progression-free 
                               survival effect estimate from the univariate
                               independent parameters (IP) model  into the surrogacy models, 
                               using data available at present-day.} 
\label{tab:predOsIpPresent}
\begin{tabular}{lrrrrr}
  \toprule
Indication & IP & CP & MCIP & RP & MRIP \\ 
  \midrule
CRC & -0.22 ( -0.36, -0.09) & -0.17 ( -0.25, -0.07) & -0.17 ( -0.26, -0.08) & -0.20 ( -0.31, -0.09) & -0.20 ( -0.31, -0.09) \\ 
  RCC & -0.07 ( -9.76, 10.42) & -0.21 ( -5.06,  4.07) & -0.20 ( -4.86,  3.96) & -0.15 ( -4.55,  4.22) & -0.15 ( -4.87,  4.57) \\ 
  BC & -0.09 ( -0.23,  0.05) & -0.12 ( -0.26,  0.00) & -0.12 ( -0.26,  0.00) & -0.10 ( -0.24,  0.03) & -0.10 ( -0.24,  0.03) \\ 
  NSCLC & -0.14 ( -0.38,  0.11) & -0.11 ( -0.26,  0.06) & -0.11 ( -0.27,  0.06) & -0.13 ( -0.30,  0.06) & -0.13 ( -0.31,  0.06) \\ 
  OFTPP & -0.08 ( -0.23,  0.09) & -0.07 ( -0.19,  0.06) & -0.07 ( -0.19,  0.06) & -0.07 ( -0.20,  0.06) & -0.07 ( -0.20,  0.06) \\ 
  CC & -0.22 (-33.56, 32.95) & -0.21 ( -7.74,  7.33) & -0.21 ( -8.22,  7.80) & -0.26 ( -9.94,  9.24) & -0.26 (-10.45,  9.71) \\ 
  GBM & -0.02 ( -0.94,  0.91) &  0.00 ( -0.62,  0.60) &  0.00 ( -0.60,  0.59) & -0.01 ( -0.63,  0.61) & -0.01 ( -0.63,  0.62) \\ 
   \bottomrule
\end{tabular}
\end{table}
\begin{table}[H]
\centering
\caption{Mean and 95\% credible interval estimates for the
                               predicted effect on overall survival (log hazard ratio scale), from entering the  progression-free 
                               survival effect estimate from each
                               univariate model into the corresponding surrogacy model, 
                               using data available at present-day.} 
\label{tab:predOsMatchedPresent}
\begin{tabular}{lrrrrr}
  \toprule
Indication & IP & CP & MCIP & RP & MRIP \\ 
  \midrule
CRC & -0.22 ( -0.36, -0.09) & -0.11 ( -0.16, -0.06) & -0.12 ( -0.19, -0.07) & -0.19 ( -0.29, -0.10) & -0.19 ( -0.29, -0.10) \\ 
  RCC & -0.07 ( -9.76, 10.42) & -0.11 ( -0.16, -0.06) & -0.12 ( -0.18, -0.06) & -0.14 ( -0.27, -0.02) & -0.14 ( -0.28, -0.01) \\ 
  BC & -0.09 ( -0.23,  0.05) & -0.11 ( -0.16, -0.06) & -0.11 ( -0.17, -0.06) & -0.09 ( -0.17, -0.01) & -0.09 ( -0.18, -0.01) \\ 
  NSCLC & -0.14 ( -0.38,  0.11) & -0.11 ( -0.16, -0.06) & -0.11 ( -0.17, -0.06) & -0.14 ( -0.25, -0.04) & -0.14 ( -0.26, -0.03) \\ 
  OFTPP & -0.08 ( -0.23,  0.09) & -0.11 ( -0.16, -0.06) & -0.11 ( -0.17, -0.05) & -0.09 ( -0.18,  0.00) & -0.09 ( -0.18,  0.01) \\ 
  CC & -0.22 (-33.56, 32.95) & -0.11 ( -0.16, -0.06) & -0.11 ( -0.24, -0.04) & -0.17 ( -0.37, -0.01) & -0.16 ( -0.40,  0.03) \\ 
  GBM & -0.02 ( -0.94,  0.91) & -0.11 ( -0.16, -0.06) & -0.12 ( -0.17, -0.04) & -0.07 ( -0.19,  0.05) & -0.07 ( -0.21,  0.06) \\ 
  All Indications &  & -0.11 ( -0.16, -0.06) & -0.12 ( -0.17, -0.06) & -0.13 ( -0.23, -0.03) & -0.13 ( -0.23, -0.03) \\ 
  New Indication &  &  &  & -0.13 ( -0.38,  0.09) & -0.13 ( -0.38,  0.10) \\ 
   \bottomrule
\end{tabular}
\end{table}
			\subsubsection{Deviance information criterion (DIC)} \label{sec:appPresentDic}
\begin{table}[H]
\centering
\caption{Residual deviance, deviance, number of effective parameters, and deviance information criterian (DIC) from the application of univariate models to data on overall survival at present-day. *Based on 38 data points.} 
\label{tab:dicOsPresent}
\begin{tabular}{lrrrr}
  \toprule
Model & Residual deviance & Deviance & $p_D$ & DIC \\ 
  \midrule
IP & 33.58 & -39.04 & 20.26 & -18.78 \\ 
  CP & 33.80 & -38.82 & 17.06 & -21.76 \\ 
  MCIP & 33.79 & -38.83 & 17.10 & -21.73 \\ 
  RP & 33.38 & -39.24 & 17.87 & -21.37 \\ 
  MRIP & 33.42 & -37.81 & 17.89 & -19.92 \\ 
   \bottomrule
\end{tabular}
\end{table}

\begin{table}[H]
\centering
\caption{Residual deviance, deviance, number of effective parameters, and deviance information criterian (DIC) from the application of univariate models to data on progression-free survival at present-day. *Based on 43 data points.} 
\label{tab:dicPfsPresent}
\begin{tabular}{lrrrr}
  \toprule
Model & Residual deviance & Deviance & $p_D$ & DIC \\ 
  \midrule
IP & 42.48 & -44.76 & 29.17 & -15.59 \\ 
  CP & 45.28 & -41.97 & 28.96 & -13.00 \\ 
  MCIP & 45.19 & -42.05 & 28.93 & -13.13 \\ 
  RP & 43.44 & -43.80 & 28.35 & -15.45 \\ 
  MRIP & 43.48 & -42.38 & 28.28 & -14.10 \\ 
   \bottomrule
\end{tabular}
\end{table}

			\clearpage
		\end{landscape}
		
		\begin{landscape}
			\subsubsection{Surrogacy parameter estimates} \label{sec:appPresentSurrogacy} 
\begin{table}[H]
\centering
\caption{Median and 95\% credible interval estimates for the intercept parameter in the 
                       surrogacy models using data available at present-day. N - number of trials per indication.} 
\label{tab:surrogateInterceptPresent}
\begin{tabular}{lrrrrrr}
  \toprule
Indication & N & IP & CP & MCIP & RP & MRIP \\ 
  \midrule
CRC &  9 &  0.05 ( -0.11,  0.21) &  0.04 ( -0.04,  0.12) &  0.04 ( -0.04,  0.13) &  0.04 ( -0.06,  0.12) &  0.03 ( -0.07,  0.13) \\ 
  RCC &  2 &  0.05 (-12.75, 15.19) &  0.04 ( -0.04,  0.12) &  0.04 ( -0.04,  0.13) &  0.05 ( -0.07,  0.16) &  0.05 ( -0.08,  0.17) \\ 
  BC & 10 &  0.04 ( -0.33,  0.45) &  0.04 ( -0.04,  0.12) &  0.04 ( -0.04,  0.13) &  0.05 ( -0.06,  0.14) &  0.04 ( -0.06,  0.14) \\ 
  NSCLC &  6 &  0.10 ( -0.18,  0.38) &  0.04 ( -0.04,  0.12) &  0.04 ( -0.04,  0.13) &  0.05 ( -0.06,  0.15) &  0.05 ( -0.06,  0.15) \\ 
  OFTPP &  7 &  0.02 ( -0.18,  0.28) &  0.04 ( -0.04,  0.12) &  0.04 ( -0.04,  0.13) &  0.05 ( -0.04,  0.15) &  0.05 ( -0.05,  0.15) \\ 
  CC &  1 & -0.40 (-22.89, 23.86) &  0.04 ( -0.04,  0.12) &  0.04 ( -0.05,  0.14) &  0.03 ( -0.12,  0.14) &  0.03 ( -0.13,  0.16) \\ 
  GBM &  3 &  0.11 ( -1.52,  1.64) &  0.04 ( -0.04,  0.12) &  0.04 ( -0.04,  0.13) &  0.07 ( -0.04,  0.21) &  0.06 ( -0.05,  0.21) \\ 
  Overall & 38 &  &  0.04 ( -0.04,  0.12) &  0.04 ( -0.04,  0.13) &  0.05 ( -0.05,  0.14) &  0.05 ( -0.05,  0.14) \\ 
   \bottomrule
\end{tabular}
\end{table}
\begin{table}[H]
\centering
\caption{Median and 95\% credible interval estimates for the slope  parameter in the 
                       surrogacy models using data available at present-day. N - number of trials per indication.} 
\label{tab:surrogateSlopePresent}
\begin{tabular}{lrrrrrr}
  \toprule
Indication & N & IP & CP & MCIP & RP & MRIP \\ 
  \midrule
CRC &  9 &  0.59 (  0.24,  0.91) &  0.42 (  0.23,  0.61) &  0.43 (  0.22,  0.66) &  0.53 (  0.29,  0.74) &  0.52 (  0.28,  0.74) \\ 
  RCC &  2 &  0.44 (-28.08, 34.37) &  0.42 (  0.23,  0.61) &  0.42 (  0.22,  0.63) &  0.43 (  0.14,  0.72) &  0.43 (  0.12,  0.73) \\ 
  BC & 10 &  0.40 ( -0.55,  1.60) &  0.42 (  0.23,  0.61) &  0.42 (  0.22,  0.63) &  0.42 (  0.15,  0.70) &  0.42 (  0.15,  0.69) \\ 
  NSCLC &  6 &  0.58 (  0.02,  1.12) &  0.42 (  0.23,  0.61) &  0.42 (  0.22,  0.63) &  0.48 (  0.23,  0.70) &  0.47 (  0.22,  0.71) \\ 
  OFTPP &  7 &  0.24 ( -0.20,  0.75) &  0.42 (  0.23,  0.61) &  0.41 (  0.21,  0.63) &  0.34 (  0.10,  0.57) &  0.34 (  0.09,  0.58) \\ 
  CC &  1 & -0.47 (-54.65, 59.38) &  0.42 (  0.23,  0.61) &  0.42 (  0.21,  0.75) &  0.49 (  0.16,  0.92) &  0.47 (  0.09,  0.98) \\ 
  GBM &  3 &  0.30 ( -3.14,  3.47) &  0.42 (  0.23,  0.61) &  0.42 (  0.20,  0.63) &  0.34 (  0.03,  0.60) &  0.35 (  0.01,  0.63) \\ 
  Overall & 38 &  &  0.42 (  0.23,  0.61) &  0.42 (  0.22,  0.63) &  0.43 (  0.20,  0.66) &  0.43 (  0.19,  0.66) \\ 
   \bottomrule
\end{tabular}
\end{table}
\begin{table}[H]
\centering
\caption{Median and 95\% credible interval estimates for the 
conditonal variance parameter in the 
                       surrogacy models using data available at present-day. N - number of trials per indication.} 
\label{tab:surrogatePsiPresent}
\begin{tabular}{lrrrrrr}
  \toprule
Indication & N & IP & CP & MCIP & RP & MRIP \\ 
  \midrule
CRC &  9 & 0.05 (0.00, 0.17) & 0.05 (0.00, 0.11) & 0.05 (0.00, 0.13) & 0.04 (0.00, 0.11) & 0.04 (0.00, 0.13) \\ 
  RCC &  2 & 0.26 (0.01, 1.02) & 0.05 (0.00, 0.11) & 0.05 (0.00, 0.26) & 0.04 (0.00, 0.15) & 0.05 (0.00, 0.33) \\ 
  BC & 10 & 0.07 (0.00, 0.26) & 0.05 (0.00, 0.11) & 0.04 (0.00, 0.14) & 0.04 (0.00, 0.13) & 0.04 (0.00, 0.16) \\ 
  NSCLC &  6 & 0.08 (0.00, 0.34) & 0.05 (0.00, 0.11) & 0.05 (0.00, 0.15) & 0.04 (0.00, 0.14) & 0.05 (0.00, 0.17) \\ 
  OFTPP &  7 & 0.06 (0.00, 0.30) & 0.05 (0.00, 0.11) & 0.05 (0.00, 0.19) & 0.04 (0.00, 0.14) & 0.04 (0.00, 0.18) \\ 
  CC &  1 & 0.34 (0.02, 1.12) & 0.05 (0.00, 0.11) & 0.07 (0.01, 0.75) & 0.05 (0.00, 0.19) & 0.08 (0.00, 0.74) \\ 
  GBM &  3 & 0.23 (0.02, 0.91) & 0.05 (0.00, 0.11) & 0.08 (0.01, 0.48) & 0.05 (0.00, 0.18) & 0.07 (0.00, 0.39) \\ 
  Overall & 38 &  & 0.05 (0.00, 0.11) & 0.05 (0.00, 0.13) &  &  \\ 
   \bottomrule
\end{tabular}
\end{table}
			\subsubsection{Deviance information criterion (DIC)} \label{sec:appPresentSurrogateDic}
\begin{table}[H]
\centering
\caption{Residual deviance, deviance, number of effective parameters, and deviance information criterian (DIC) from the application of surrogacy models to data at present-day. *Based on 38 data points.} 
\label{tab:dicSurrogatePresent}
\begin{tabular}{lrrrr}
  \toprule
Model & Residual deviance & Deviance & $p_D$ & DIC \\ 
  \midrule
IP & 70.95 & -127.59 & 48.75 & -78.84 \\ 
  CP & 72.75 & -118.81 & 39.52 & -79.29 \\ 
  MCIP & 72.35 & -115.95 & 41.27 & -74.68 \\ 
  RP & 71.92 & -122.36 & 42.80 & -79.56 \\ 
  MRIP & 71.61 & -114.92 & 43.01 & -71.91 \\ 
   \bottomrule
\end{tabular}
\end{table}

			\subsubsection{Mixture probabilities} \label{sec:appPresentSurrogateMixProbs}
\begin{table}[H]
\centering
\caption{Mixture probability estimates (as mean and standard deviation), and informative and vague mixture component estimates (as median and 95\% credible interval), for the intercept parameter from the application of surrogacy models to data at present-day. } 
\label{tab:mixProbsInterceptsPresentWithBranch}
\begin{tabular}{llrrr}
  \toprule
Model & Indication & Probability & Informative & Vague \\ 
  \midrule
MCIP & BC & 1.00 (0.04) & 0.04 (-0.04, 0.13) &  0.14 (-61.89, 61.86) \\ 
  MCIP & CC & 0.98 (0.14) & 0.04 (-0.04, 0.13) &  0.04 (-61.76, 61.83) \\ 
  MCIP & CRC & 0.99 (0.07) & 0.04 (-0.04, 0.13) & -0.03 (-62.00, 61.82) \\ 
  MCIP & GBM & 0.99 (0.11) & 0.04 (-0.04, 0.13) &  0.13 (-61.82, 62.14) \\ 
  MCIP & NSCLC & 1.00 (0.04) & 0.04 (-0.04, 0.13) & -0.08 (-61.91, 61.88) \\ 
  MCIP & OFTPP & 1.00 (0.05) & 0.04 (-0.04, 0.13) &  0.11 (-62.11, 62.06) \\ 
  MCIP & RCC & 1.00 (0.05) & 0.04 (-0.04, 0.13) &  0.11 (-62.02, 62.11) \\ 
   &  &  &  &  \\ 
  MRIP & BC & 1.00 (0.04) & 0.04 (-0.06, 0.14) &  0.04 (-61.87, 61.87) \\ 
  MRIP & CC & 0.99 (0.10) & 0.03 (-0.12, 0.16) & -0.07 (-62.06, 61.74) \\ 
  MRIP & CRC & 1.00 (0.05) & 0.03 (-0.07, 0.13) &  0.02 (-61.95, 61.82) \\ 
  MRIP & GBM & 0.99 (0.09) & 0.06 (-0.05, 0.21) &  0.10 (-61.85, 61.38) \\ 
  MRIP & NSCLC & 1.00 (0.06) & 0.05 (-0.06, 0.15) &  0.02 (-61.94, 62.08) \\ 
  MRIP & OFTPP & 1.00 (0.06) & 0.05 (-0.05, 0.15) &  0.09 (-61.78, 61.94) \\ 
  MRIP & RCC & 1.00 (0.06) & 0.05 (-0.08, 0.17) &  0.02 (-62.08, 61.97) \\ 
   \bottomrule
\end{tabular}
\end{table}
\begin{table}[H]
\centering
\caption{Mixture probability estimates (as mean and standard deviation), and informative and vague mixture component estimates (as median and 95\% credible interval), for the slope parameter from the application of surrogacy models to data at present-day. } 
\label{tab:mixProbsSlopesPresentWithBranch}
\begin{tabular}{llrrr}
  \toprule
Model & Indication & Probability & Informative & Vague \\ 
  \midrule
MCIP & BC & 1.00 (0.06) & 0.42 ( 0.22, 0.63) &  0.17 (-62.07, 62.29) \\ 
  MCIP & CC & 0.96 (0.20) & 0.42 ( 0.22, 0.63) &  0.61 (-61.25, 61.56) \\ 
  MCIP & CRC & 0.92 (0.26) & 0.42 ( 0.22, 0.63) &  0.54 (-60.83, 60.93) \\ 
  MCIP & GBM & 0.98 (0.14) & 0.42 ( 0.22, 0.63) &  0.18 (-61.74, 61.74) \\ 
  MCIP & NSCLC & 0.99 (0.08) & 0.42 ( 0.22, 0.63) &  0.32 (-61.88, 61.93) \\ 
  MCIP & OFTPP & 0.98 (0.15) & 0.42 ( 0.22, 0.63) &  0.23 (-61.61, 61.29) \\ 
  MCIP & RCC & 1.00 (0.07) & 0.42 ( 0.22, 0.63) &  0.17 (-62.14, 62.01) \\ 
   &  &  &  &  \\ 
  MRIP & BC & 0.99 (0.08) & 0.42 ( 0.15, 0.69) &  0.26 (-61.93, 62.03) \\ 
  MRIP & CC & 0.97 (0.17) & 0.47 ( 0.11, 0.91) &  0.53 (-61.42, 61.47) \\ 
  MRIP & CRC & 0.98 (0.14) & 0.52 ( 0.27, 0.74) &  0.47 (-61.77, 61.85) \\ 
  MRIP & GBM & 0.98 (0.12) & 0.35 ( 0.02, 0.63) &  0.15 (-61.60, 61.96) \\ 
  MRIP & NSCLC & 0.99 (0.09) & 0.47 ( 0.22, 0.71) &  0.21 (-62.07, 61.96) \\ 
  MRIP & OFTPP & 0.99 (0.12) & 0.34 ( 0.10, 0.59) &  0.20 (-61.65, 61.76) \\ 
  MRIP & RCC & 0.99 (0.10) & 0.43 ( 0.12, 0.73) &  0.22 (-61.70, 61.56) \\ 
   \bottomrule
\end{tabular}
\end{table}
\begin{table}[H]
\centering
\caption{Mixture probability estimates (as mean and standard deviation), and informative and vague mixture component estimates (as median and 95\% credible interval), for the conditional variance parameter from the application of surrogacy models to data at present-day. } 
\label{tab:mixProbsVariancesPresentWithBranch}
\begin{tabular}{llrrr}
  \toprule
Model & Indication & Probability & Informative & Vague \\ 
  \midrule
MCIP & BC & 0.82 (0.38) & 0.05 ( 0.00, 0.13) &  0.26 (  0.01,  1.08) \\ 
  MCIP & CC & 0.58 (0.49) & 0.05 ( 0.00, 0.13) &  0.28 (  0.01,  1.06) \\ 
  MCIP & CRC & 0.82 (0.38) & 0.05 ( 0.00, 0.13) &  0.25 (  0.01,  1.08) \\ 
  MCIP & GBM & 0.57 (0.50) & 0.05 ( 0.00, 0.13) &  0.24 (  0.02,  1.02) \\ 
  MCIP & NSCLC & 0.81 (0.39) & 0.05 ( 0.00, 0.13) &  0.25 (  0.01,  1.08) \\ 
  MCIP & OFTPP & 0.78 (0.41) & 0.05 ( 0.00, 0.13) &  0.25 (  0.01,  1.07) \\ 
  MCIP & RCC & 0.78 (0.41) & 0.05 ( 0.00, 0.13) &  0.26 (  0.01,  1.07) \\ 
   &  &  &  &  \\ 
  MRIP & BC & 0.79 (0.40) & 0.04 ( 0.00, 0.16) &  0.24 (  0.01,  1.08) \\ 
  MRIP & CC & 0.60 (0.49) & 0.05 ( 0.00, 0.22) &  0.29 (  0.01,  1.06) \\ 
  MRIP & CRC & 0.82 (0.39) & 0.04 ( 0.00, 0.15) &  0.25 (  0.01,  1.08) \\ 
  MRIP & GBM & 0.68 (0.46) & 0.06 ( 0.00, 0.21) &  0.24 (  0.01,  1.05) \\ 
  MRIP & NSCLC & 0.79 (0.41) & 0.04 ( 0.00, 0.17) &  0.24 (  0.01,  1.07) \\ 
  MRIP & OFTPP & 0.80 (0.40) & 0.04 ( 0.00, 0.17) &  0.25 (  0.01,  1.07) \\ 
  MRIP & RCC & 0.75 (0.43) & 0.04 ( 0.00, 0.19) &  0.25 (  0.01,  1.06) \\ 
   \bottomrule
\end{tabular}
\end{table}
			\clearpage
		\end{landscape}
		
		\begin{landscape}
		\subsection{TA178 (renal)}
			\subsubsection{Pooled OS and PFS estimates}  \label{sec:appTa178Pooled}
\begin{table}[H]
\centering
\caption{Median and 95\% credible interval estimates from the application of univariate models to effects on overall 
                       survival (log hazard ratio scale) using data available at the time of NICE TA178.} 
\label{tab:os_ta178}
\begin{tabular}{lrrrrr}
  \toprule
Indication & IP & CP & MCIP & RP & MRIP \\ 
  \midrule
CRC & -0.23 (-0.41, -0.06) & -0.16 (-0.25, -0.06) & -0.16 (-0.25, -0.06) & -0.19 (-0.33, -0.06) & -0.19 (-0.33, -0.06) \\ 
  BC & -0.10 (-0.29,  0.10) & -0.16 (-0.25, -0.06) & -0.16 (-0.25, -0.06) & -0.13 (-0.26,  0.02) & -0.13 (-0.26,  0.01) \\ 
  NSCLC & -0.11 (-0.73,  0.52) & -0.16 (-0.25, -0.06) & -0.16 (-0.25, -0.06) & -0.14 (-0.37,  0.10) & -0.14 (-0.35,  0.08) \\ 
  RCC & -0.15 (-0.65,  0.35) & -0.16 (-0.25, -0.06) & -0.16 (-0.25, -0.06) & -0.15 (-0.34,  0.04) & -0.15 (-0.33,  0.03) \\ 
  All Indications &  & -0.16 (-0.25, -0.06) & -0.16 (-0.25, -0.06) & -0.16 (-0.38,  0.08) & -0.15 (-0.34,  0.04) \\ 
  New Indication &  &  &  & -0.16 (-0.61,  0.31) & -0.15 (-0.52,  0.23) \\ 
   \bottomrule
\end{tabular}
\end{table}
\begin{table}[H]
\centering
\caption{Median and 95\% credible interval estimates from the application of univariate models to effects on progression-free 
                       survival (log hazard ratio scale) using data available at the time of NICE TA178.} 
\label{tab:pfs_ta178}
\begin{tabular}{lrrrrr}
  \toprule
Indication & IP & CP & MCIP & RP & MRIP \\ 
  \midrule
CRC & -0.48 (-0.76, -0.26) & -0.40 (-0.52, -0.30) & -0.40 (-0.52, -0.29) & -0.43 (-0.65, -0.28) & -0.43 (-0.64, -0.28) \\ 
  BC & -0.34 (-0.54, -0.15) & -0.40 (-0.52, -0.30) & -0.40 (-0.52, -0.29) & -0.37 (-0.53, -0.22) & -0.37 (-0.53, -0.22) \\ 
  NSCLC & -0.29 (-0.91,  0.33) & -0.40 (-0.52, -0.30) & -0.40 (-0.52, -0.29) & -0.37 (-0.63, -0.09) & -0.37 (-0.62, -0.11) \\ 
  RCC & -0.50 (-1.01, -0.15) & -0.40 (-0.52, -0.30) & -0.40 (-0.52, -0.29) & -0.44 (-0.70, -0.25) & -0.44 (-0.68, -0.26) \\ 
  All Indications &  & -0.40 (-0.52, -0.30) & -0.40 (-0.52, -0.29) & -0.40 (-0.69, -0.13) & -0.40 (-0.64, -0.18) \\ 
  New Indication &  &  &  & -0.40 (-0.98,  0.17) & -0.40 (-0.85,  0.04) \\ 
   \bottomrule
\end{tabular}
\end{table}
			\subsubsection{Predicted OS estimates}
\begin{table}[H]
\centering
\caption{Median and 95\% credible interval estimates for 
                               predicted effects on overall survival (log hazard ratio scale), from entering the progression-free 
                               survival effect estimate from the univariate
                               independent parameters (IP) model into the surrogacy models, 
                               using data available at the time of NICE TA178.} 
\label{tab:predOsIp_ta178}
\begin{tabular}{lrrrrr}
  \toprule
Indication & IP & CP & MCIP & RP & MRIP \\ 
  \midrule
RCC & -0.16 ( -8.63,  8.18) & -0.15 (-14.79, 14.26) & -0.16 (-16.27, 15.96) & -0.14 (-13.51, 13.31) & -0.15 (-15.45, 15.08) \\ 
   \bottomrule
\end{tabular}
\end{table}

\begin{table}[H]
\centering
\caption{Median and 95\% credible interval estimates for 
                               predicted effects on overall survival (log hazard ratio scale), from entering the progression-free 
                               survival effect estimate from each univariate
                                model into the corresponding surrogacy model, 
                               using data available at the time of NICE TA178.} 
\label{tab:predOsMatched_ta178}
\begin{tabular}{lrrrrr}
  \toprule
Indication & IP & CP & MCIP & RP & MRIP \\ 
  \midrule
RCC & -0.16 (-8.63, 8.18) & -0.16 (-0.59, 0.28) & -0.17 (-0.68, 0.30) & -0.12 (-0.51, 0.28) & -0.14 (-0.64, 0.32) \\ 
   \bottomrule
\end{tabular}
\end{table}

			\subsubsection{Deviance information criterion (DIC)} \label{sec:appTa178Dic}
\begin{table}[H]
\centering
\caption{Residual deviance, deviance, number of effective parameters, and deviance information criterian (DIC) from the application of univariate models to data on overall survival at the time of NICE TA178. *Based on 13 data points.} 
\label{tab:dicOs_ta178}
\begin{tabular}{lrrrr}
  \toprule
Model & Residual deviance & Deviance & $p_D$ & DIC \\ 
  \midrule
IP & 11.13 & -16.92 & 8.99 & -7.92 \\ 
  CP & 10.92 & -17.13 & 7.56 & -9.57 \\ 
  MCIP & 10.94 & -17.11 & 7.59 & -9.52 \\ 
  RP & 10.87 & -17.18 & 8.23 & -8.95 \\ 
  MRIP & 10.84 & -15.82 & 8.16 & -7.66 \\ 
   \bottomrule
\end{tabular}
\end{table}

\begin{table}[H]
\centering
\caption{Residual deviance, deviance, number of effective parameters, and deviance information criterian (DIC) from the application of univariate models to data on progression-free survival at the time of NICE TA178. *Based on 17 data points.} 
\label{tab:dicPfs_ta178}
\begin{tabular}{lrrrr}
  \toprule
Model & Residual deviance & Deviance & $p_D$ & DIC \\ 
  \midrule
IP & 18.30 & -15.98 & 13.56 & -2.42 \\ 
  CP & 19.48 & -14.80 & 12.37 & -2.43 \\ 
  MCIP & 19.50 & -14.78 & 12.42 & -2.36 \\ 
  RP & 18.81 & -15.47 & 12.91 & -2.56 \\ 
  MRIP & 18.87 & -14.02 & 12.86 & -1.16 \\ 
   \bottomrule
\end{tabular}
\end{table}

			\subsubsection{Mixture probabilities} \label{sec:appTa178MixProbs}
\begin{table}[H]
\centering
\caption{Mixture probability estimates (as mean and standard deviation), and informative and vague mixture component estimates (as median and 95\% credible interval), from the application of univariate models to data on overall survival at the time of NICE TA178.} 
\label{tab:mixProbsOsWithBranch_ta178}
\begin{tabular}{llrrr}
  \toprule
Model & Indication & Probability & Informative & Vague \\ 
  \midrule
MCIP & BC & 0.99 (0.07) & -0.16 (-0.25, -0.06) & -0.08 (-61.86, 61.60) \\ 
  MCIP & CRC & 0.99 (0.09) & -0.16 (-0.25, -0.06) & -0.12 (-61.93, 62.00) \\ 
  MCIP & NSCLC & 0.99 (0.08) & -0.16 (-0.25, -0.06) & -0.03 (-62.07, 61.98) \\ 
  MCIP & RCC & 0.99 (0.07) & -0.16 (-0.25, -0.06) & -0.02 (-61.95, 61.66) \\ 
 & & & & \\
  MRIP & BC & 0.99 (0.08) & -0.13 (-0.26,  0.01) & -0.08 (-61.97, 61.69) \\ 
  MRIP & CRC & 0.99 (0.08) & -0.19 (-0.33, -0.05) & -0.13 (-61.90, 61.54) \\ 
  MRIP & NSCLC & 0.99 (0.09) & -0.14 (-0.35,  0.08) & -0.05 (-61.84, 61.86) \\ 
  MRIP & RCC & 0.99 (0.09) & -0.15 (-0.34,  0.03) & -0.07 (-62.01, 62.01) \\ 
   \bottomrule
\end{tabular}
\end{table}
\begin{table}[H]
\centering
\caption{Mixture probability estimates (as mean and standard deviation), and informative and vague mixture component estimates (as median and 95\% credible interval), from the application of univariate models to data on progression-free survival at the time of NICE TA178.} 
\label{tab:mixProbsPfsWithBranch_ta178}
\begin{tabular}{llrrr}
  \toprule
Model & Indication & Probability & Informative & Vague \\ 
  \midrule
MCIP & BC & 1.00 (0.07) & -0.40 (-0.52, -0.29) & -0.22 (-62.13, 61.50) \\ 
  MCIP & CRC & 0.99 (0.08) & -0.40 (-0.52, -0.29) & -0.24 (-62.03, 62.10) \\ 
  MCIP & NSCLC & 0.99 (0.10) & -0.40 (-0.52, -0.29) & -0.15 (-61.99, 61.80) \\ 
  MCIP & RCC & 0.99 (0.09) & -0.40 (-0.52, -0.29) & -0.27 (-62.29, 61.98) \\
 & & & & \\ 
  MRIP & BC & 0.99 (0.09) & -0.37 (-0.53, -0.22) & -0.14 (-61.44, 61.65) \\ 
  MRIP & CRC & 0.99 (0.08) & -0.43 (-0.64, -0.28) & -0.15 (-62.20, 61.48) \\ 
  MRIP & NSCLC & 0.99 (0.11) & -0.37 (-0.62, -0.11) & -0.12 (-62.08, 61.79) \\ 
  MRIP & RCC & 0.99 (0.10) & -0.44 (-0.68, -0.25) & -0.41 (-61.91, 62.15) \\ 
   \bottomrule
\end{tabular}
\end{table}
			\clearpage
		\end{landscape}
		
		\begin{landscape}
		\subsection{TA285 (ovarian)}
			\subsubsection{Pooled OS and PFS estimates} \label{sec:appTa285Pooled}
\begin{table}[H]
\centering
\caption{Median and 95\% credible interval estimates from the application of univariate models to effects on overall 
                       survival (log hazard ratio scale) using data available at the time of NICE TA285.} 
\label{tab:os_ta285}
\begin{tabular}{lrrrrr}
  \toprule
Indication & IP & CP & MCIP & RP & MRIP \\ 
  \midrule
CRC & -0.19 (-0.33, -0.07) & -0.13 (-0.20, -0.07) & -0.13 (-0.20, -0.07) & -0.16 (-0.26, -0.07) & -0.16 (-0.26, -0.07) \\ 
  RCC & -0.15 (-0.65,  0.34) & -0.13 (-0.20, -0.07) & -0.13 (-0.20, -0.07) & -0.14 (-0.27, -0.01) & -0.14 (-0.27, -0.01) \\ 
  BC & -0.12 (-0.31,  0.04) & -0.13 (-0.20, -0.07) & -0.13 (-0.20, -0.07) & -0.13 (-0.23, -0.02) & -0.13 (-0.23, -0.03) \\ 
  NSCLC & -0.18 (-0.63,  0.25) & -0.13 (-0.20, -0.07) & -0.13 (-0.20, -0.07) & -0.14 (-0.29, -0.01) & -0.14 (-0.29, -0.01) \\ 
  CC & -0.34 (-1.46,  0.78) & -0.13 (-0.20, -0.07) & -0.13 (-0.21, -0.07) & -0.14 (-0.36,  0.02) & -0.14 (-0.37,  0.02) \\ 
  GBM & -0.02 (-0.62,  0.61) & -0.13 (-0.20, -0.07) & -0.13 (-0.20, -0.07) & -0.12 (-0.26,  0.06) & -0.12 (-0.26,  0.06) \\ 
  OFTPP & -0.06 (-0.25,  0.12) & -0.13 (-0.20, -0.07) & -0.13 (-0.20, -0.07) & -0.11 (-0.21,  0.00) & -0.11 (-0.21,  0.01) \\ 
  All Indications &  & -0.13 (-0.20, -0.07) & -0.13 (-0.20, -0.07) & -0.13 (-0.24, -0.04) & -0.13 (-0.23, -0.04) \\ 
  New Indication &  &  &  & -0.13 (-0.34,  0.07) & -0.13 (-0.34,  0.06) \\ 
   \bottomrule
\end{tabular}
\end{table}
\begin{table}[H]
\centering
\caption{Median and 95\% credible interval estimates from the application of univariate models to effects on progression-free 
                       survival (log hazard ratio scale) using data available at the time of NICE TA285.} 
\label{tab:pfs_ta285}
\begin{tabular}{lrrrrr}
  \toprule
Indication & IP & CP & MCIP & RP & MRIP \\ 
  \midrule
CRC & -0.47 (-0.67, -0.30) & -0.38 (-0.47, -0.31) & -0.38 (-0.48, -0.30) & -0.42 (-0.58, -0.30) & -0.42 (-0.58, -0.30) \\ 
  RCC & -0.50 (-1.01, -0.15) & -0.38 (-0.47, -0.31) & -0.38 (-0.48, -0.30) & -0.42 (-0.62, -0.28) & -0.42 (-0.62, -0.28) \\ 
  BC & -0.30 (-0.42, -0.18) & -0.38 (-0.47, -0.31) & -0.38 (-0.47, -0.30) & -0.34 (-0.45, -0.23) & -0.34 (-0.45, -0.23) \\ 
  NSCLC & -0.47 (-1.08,  0.11) & -0.38 (-0.47, -0.31) & -0.38 (-0.48, -0.30) & -0.40 (-0.63, -0.22) & -0.40 (-0.63, -0.22) \\ 
  CC & -0.40 (-1.50,  0.69) & -0.38 (-0.47, -0.31) & -0.38 (-0.48, -0.30) & -0.39 (-0.61, -0.19) & -0.39 (-0.62, -0.19) \\ 
  GBM & -0.35 (-0.93,  0.24) & -0.38 (-0.47, -0.31) & -0.38 (-0.48, -0.30) & -0.38 (-0.56, -0.21) & -0.38 (-0.56, -0.21) \\ 
  OFTPP & -0.43 (-0.90,  0.02) & -0.38 (-0.47, -0.31) & -0.38 (-0.48, -0.30) & -0.40 (-0.61, -0.22) & -0.40 (-0.61, -0.23) \\ 
  All Indications &  & -0.38 (-0.47, -0.31) & -0.38 (-0.48, -0.30) & -0.39 (-0.54, -0.28) & -0.39 (-0.53, -0.28) \\ 
  New Indication &  &  &  & -0.39 (-0.68, -0.14) & -0.39 (-0.67, -0.15) \\ 
   \bottomrule
\end{tabular}
\end{table}
			\subsubsection{Predicted OS estimates}
\begin{table}[H]
\centering
\caption{Median and 95\% credible interval estimates for 
                               predicted effects on overall survival (log hazard ratio scale), from entering the progression-free 
                               survival effect estimate from the univariate
                               independent parameters (IP) model into the surrogacy models, 
                               using data available at the time of NICE TA285.} 
\label{tab:predOsIp_ta285}
\begin{tabular}{lrrrrr}
  \toprule
Indication & IP & CP & MCIP & RP & MRIP \\ 
  \midrule
OFTPP & -0.07 (-0.39, 0.27) & -0.07 (-0.48, 0.37) & -0.07 (-0.49, 0.38) & -0.06 (-0.46, 0.35) & -0.06 (-0.45, 0.33) \\ 
   \bottomrule
\end{tabular}
\end{table}

\begin{table}[H]
\centering
\caption{Median and 95\% credible interval estimates for 
                               predicted effects on overall survival (log hazard ratio scale), from entering the progression-free 
                               survival effect estimate from each univariate
                               model into the corresponding surrogacy model, 
                               using data available at the time of NICE TA285.} 
\label{tab:predOsMatched_ta285}
\begin{tabular}{lrrrrr}
  \toprule
Indication & IP & CP & MCIP & RP & MRIP \\ 
  \midrule
OFTPP & -0.07 (-0.39, 0.27) & -0.13 (-0.47, 0.21) & -0.13 (-0.51, 0.23) & -0.06 (-0.33, 0.18) & -0.05 (-0.31, 0.18) \\ 
   \bottomrule
\end{tabular}
\end{table}

			\subsubsection{Deviance information criterion (DIC)} \label{sec:appTa285Dic}
\begin{table}[H]
\centering
\caption{Residual deviance, deviance, number of effective parameters, and deviance information criterian (DIC) from the application of univariate models to data on overall survival at the time of NICE TA285. *Based on 27 data points.} 
\label{tab:dicOs_ta285}
\begin{tabular}{lrrrr}
  \toprule
Model & Residual deviance & Deviance & $p_D$ & DIC \\ 
  \midrule
IP & 26.94 & -30.43 & 18.83 & -11.59 \\ 
  CP & 27.60 & -29.77 & 16.02 & -13.74 \\ 
  MCIP & 27.57 & -29.80 & 15.99 & -13.81 \\ 
  RP & 27.12 & -30.25 & 16.78 & -13.47 \\ 
  MRIP & 27.12 & -28.86 & 16.75 & -12.11 \\ 
   \bottomrule
\end{tabular}
\end{table}

\begin{table}[H]
\centering
\caption{Residual deviance, deviance, number of effective parameters, and deviance information criterian (DIC) from the application of univariate models to data on progression-free survival at the time of NICE TA285. *Based on 32 data points.} 
\label{tab:dicPfs_ta285}
\begin{tabular}{lrrrr}
  \toprule
Model & Residual deviance & Deviance & $p_D$ & DIC \\ 
  \midrule
IP & 33.27 & -36.97 & 24.75 & -12.22 \\ 
  CP & 34.90 & -35.33 & 23.90 & -11.43 \\ 
  MCIP & 35.03 & -35.21 & 23.89 & -11.31 \\ 
  RP & 34.13 & -36.11 & 23.86 & -12.25 \\ 
  MRIP & 34.10 & -34.75 & 23.87 & -10.88 \\ 
   \bottomrule
\end{tabular}
\end{table}

			\subsubsection{Mixture probabilities} \label{sec:appTa285MixProbs}
\begin{table}[H]
\centering
\caption{Mixture probability estimates (as mean and standard deviation), and informative and vague mixture component estimates (as median and 95\% credible interval), from the application of univariate models to data on overall survival at the time of NICE TA285.} 
\label{tab:mixProbsOsWithBranch_ta285}
\begin{tabular}{llrrr}
  \toprule
Model & Indication & Probability & Informative & Vague \\ 
  \midrule
MCIP & BC & 1.00 (0.06) & -0.13 (-0.20, -0.07) & -0.04 (-62.02, 61.96) \\ 
  MCIP & CC & 0.99 (0.12) & -0.13 (-0.20, -0.07) & -0.26 (-61.67, 61.68) \\ 
  MCIP & CRC & 0.99 (0.07) & -0.13 (-0.20, -0.07) & -0.10 (-61.92, 62.13) \\ 
  MCIP & GBM & 0.99 (0.10) & -0.13 (-0.20, -0.07) & -0.03 (-61.37, 62.03) \\ 
  MCIP & NSCLC & 0.99 (0.07) & -0.13 (-0.20, -0.07) &  0.03 (-62.02, 61.99) \\ 
  MCIP & OFTPP & 0.99 (0.08) & -0.13 (-0.20, -0.07) & -0.04 (-61.74, 61.76) \\ 
  MCIP & RCC & 0.99 (0.07) & -0.13 (-0.20, -0.07) & -0.04 (-61.78, 61.72) \\ 
 & & & & \\
  MRIP & BC & 1.00 (0.06) & -0.13 (-0.23, -0.03) & -0.11 (-62.19, 61.87) \\ 
  MRIP & CC & 0.99 (0.12) & -0.14 (-0.35,  0.01) & -0.21 (-61.91, 61.77) \\ 
  MRIP & CRC & 0.99 (0.07) & -0.15 (-0.26, -0.07) & -0.12 (-61.81, 62.33) \\ 
  MRIP & GBM & 0.99 (0.10) & -0.12 (-0.26,  0.05) & -0.01 (-61.88, 61.65) \\ 
  MRIP & NSCLC & 1.00 (0.07) & -0.14 (-0.29, -0.01) & -0.16 (-62.12, 61.89) \\ 
  MRIP & OFTPP & 0.99 (0.08) & -0.11 (-0.22,  0.00) & -0.07 (-62.13, 62.00) \\ 
  MRIP & RCC & 1.00 (0.07) & -0.14 (-0.27, -0.01) & -0.12 (-61.84, 61.97) \\ 
   \bottomrule
\end{tabular}
\end{table}
\begin{table}[H]
\centering
\caption{Mixture probability estimates (as mean and standard deviation), and informative and vague mixture component estimates (as median and 95\% credible interval), from the application of univariate models to data on progression-free survival at the time of NICE TA285.} 
\label{tab:mixProbsPfsWithBranch_ta285}
\begin{tabular}{llrrr}
  \toprule
Model & Indication & Probability & Informative & Vague \\ 
  \midrule
MCIP & BC & 0.98 (0.13) & -0.38 (-0.48, -0.30) & -0.28 (-61.64, 61.64) \\ 
  MCIP & CC & 0.99 (0.10) & -0.38 (-0.48, -0.30) & -0.06 (-61.88, 62.08) \\ 
  MCIP & CRC & 0.99 (0.08) & -0.38 (-0.48, -0.30) & -0.21 (-62.00, 61.83) \\ 
  MCIP & GBM & 0.99 (0.07) & -0.38 (-0.48, -0.30) & -0.20 (-61.44, 62.06) \\ 
  MCIP & NSCLC & 0.99 (0.08) & -0.38 (-0.48, -0.30) & -0.18 (-61.85, 61.95) \\ 
  MCIP & OFTPP & 0.99 (0.08) & -0.38 (-0.48, -0.30) & -0.15 (-61.77, 61.92) \\ 
  MCIP & RCC & 0.99 (0.10) & -0.38 (-0.48, -0.30) & -0.42 (-61.98, 62.03) \\ 
 & & & & \\
  MRIP & BC & 0.99 (0.08) & -0.34 (-0.46, -0.23) & -0.15 (-62.29, 61.54) \\ 
  MRIP & CC & 0.99 (0.10) & -0.39 (-0.61, -0.20) & -0.28 (-61.98, 62.06) \\ 
  MRIP & CRC & 0.99 (0.08) & -0.42 (-0.57, -0.30) & -0.18 (-61.84, 62.22) \\ 
  MRIP & GBM & 0.99 (0.08) & -0.38 (-0.56, -0.21) & -0.23 (-62.05, 61.88) \\ 
  MRIP & NSCLC & 0.99 (0.09) & -0.40 (-0.63, -0.22) & -0.24 (-61.90, 61.95) \\ 
  MRIP & OFTPP & 0.99 (0.08) & -0.40 (-0.60, -0.23) & -0.23 (-61.96, 62.09) \\ 
  MRIP & RCC & 0.99 (0.09) & -0.42 (-0.61, -0.28) & -0.25 (-61.72, 62.13) \\ 
   \bottomrule
\end{tabular}
\end{table}
			\clearpage
		\end{landscape}
		
		\begin{landscape}
			\subsection{Surrogate endpoint cross validation} \label{sec:appPresentCrossVal}
\begingroup\fontsize{9pt}{10pt}\selectfont
\begin{longtable}{lrrrrrr}
\caption{Median and 95\% credible interval estimates for the treatment effect on overall
                       survival (log hazard ratio scale) from applying leave-one-out cross validation to
                       evaluate surrogacy using trial-level data at present day.} \\ 
  \toprule
Trial & Observed & IP & CP & MCIP & RP & MRIP \\ 
  \midrule
\textbf{Breast} &  &  &  &  &  &  \\ 
  E2100 & -0.13 (-0.30,  0.05) & -0.22 ( -0.75,  0.30) & -0.18 (-0.43,  0.08) & -0.18 ( -0.45,  0.08) & -0.18 (-0.47, 0.10) & -0.19 ( -0.48,  0.11) \\ 
  RIBBON-1 & -0.16 (-0.46,  0.13) & -0.12 ( -0.55,  0.32) & -0.11 (-0.44,  0.22) & -0.12 ( -0.45,  0.22) & -0.11 (-0.45, 0.23) & -0.11 ( -0.46,  0.24) \\ 
  RIBBON-1 &  0.03 (-0.26,  0.32) & -0.12 ( -0.55,  0.32) & -0.11 (-0.44,  0.22) & -0.12 ( -0.45,  0.22) & -0.11 (-0.45, 0.23) & -0.11 ( -0.46,  0.24) \\ 
  RIBBON-2 & -0.11 (-0.34,  0.13) & -0.06 ( -0.48,  0.37) & -0.06 (-0.34,  0.22) & -0.06 ( -0.35,  0.22) & -0.06 (-0.34, 0.23) & -0.06 ( -0.36,  0.24) \\ 
  AVADO &  0.03 (-0.29,  0.35) & -0.09 ( -0.54,  0.36) & -0.07 (-0.42,  0.28) & -0.07 ( -0.43,  0.29) & -0.07 (-0.43, 0.29) & -0.07 ( -0.44,  0.30) \\
\end{longtable}
\endgroup
			\clearpage
		\end{landscape}

            \subsection{Sensitivity analysis: equal mixture weights for surrogacy parameters}
                \subsubsection{Surrogacy parameter estimates} \label{sec:appPresentSurrogacyEqualMixProb}
                \begin{figure}[!htb]
				\centering
				\includegraphics[width=\textwidth]{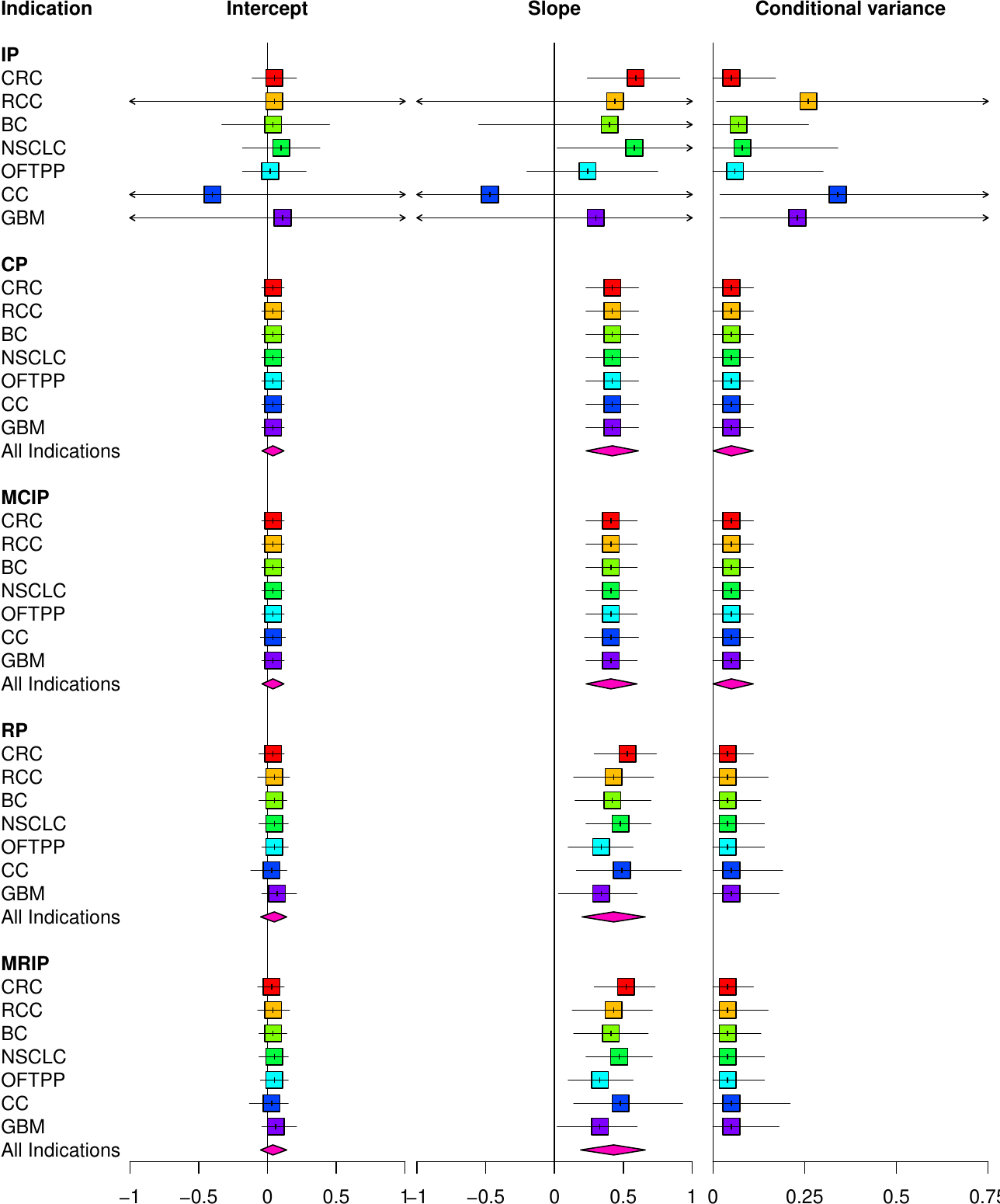}
				\caption{Multi-indication meta-analysis assuming equal mixture weights for surrogacy parameters using data available at present day. IP - independent parameters, CP - common parameter, MCIP - mixed common and independent parameters, RP - random parameters, MRIP - mixed random and independent parameters, CRC - colorectal cancer, BC - breast cancer, NSCLC - non-small cell lung cancer, RCC - renal cell carcinoma, CC - cervical cancer, GBM - glioblastoma, OFTPP - ovarian, fallopian tube, and primary peritoneal.}
				\label{fig:presentDaySurrogacyEqualMixProb}
			\end{figure}
			\clearpage

             \begin{landscape}
\begin{table}[H]
\centering
\caption{Median and 95\% credible interval estimates for the intercept parameter
                       in the surrogacy models, where mixture models assume 
                       equal mixture probabilities, using data available at present-day.} 
\label{tab:surrogateInterceptPresentEqualMixProb}
\begin{tabular}{lrrrrr}
  \toprule
Indication & IP & CP & MCIP & RP & MRIP \\ 
  \midrule
CRC &  0.05 ( -0.11,  0.21) &  0.04 ( -0.04,  0.12) &  0.04 ( -0.04,  0.12) &  0.04 ( -0.06,  0.12) &  0.03 ( -0.07,  0.12) \\ 
  RCC &  0.05 (-12.75, 15.19) &  0.04 ( -0.04,  0.12) &  0.04 ( -0.04,  0.12) &  0.05 ( -0.07,  0.16) &  0.04 ( -0.07,  0.16) \\ 
  BC &  0.04 ( -0.33,  0.45) &  0.04 ( -0.04,  0.12) &  0.04 ( -0.04,  0.12) &  0.05 ( -0.06,  0.14) &  0.04 ( -0.06,  0.14) \\ 
  NSCLC &  0.10 ( -0.18,  0.38) &  0.04 ( -0.04,  0.12) &  0.04 ( -0.04,  0.12) &  0.05 ( -0.06,  0.15) &  0.05 ( -0.06,  0.15) \\ 
  OFTPP &  0.02 ( -0.18,  0.28) &  0.04 ( -0.04,  0.12) &  0.04 ( -0.04,  0.12) &  0.05 ( -0.04,  0.15) &  0.05 ( -0.05,  0.15) \\ 
  CC & -0.40 (-22.89, 23.86) &  0.04 ( -0.04,  0.12) &  0.04 ( -0.05,  0.13) &  0.03 ( -0.12,  0.14) &  0.03 ( -0.13,  0.15) \\ 
  GBM &  0.11 ( -1.52,  1.64) &  0.04 ( -0.04,  0.12) &  0.04 ( -0.04,  0.12) &  0.07 ( -0.04,  0.21) &  0.06 ( -0.04,  0.21) \\ 
  All Indications &  &  0.04 ( -0.04,  0.12) &  0.04 ( -0.04,  0.12) &  0.05 ( -0.05,  0.14) &  0.04 ( -0.05,  0.14) \\ 
   \bottomrule
\end{tabular}
\end{table}
\begin{table}[H]
\centering
\caption{Median and 95\% credible interval estimates for the slope parameter
                       in the surrogacy models, where mixture models assume 
                       equal mixture probabilities, using data available at present-day.} 
\label{tab:surrogateSlopePresentEqualMixProb}
\begin{tabular}{lrrrrr}
  \toprule
Indication & IP & CP & MCIP & RP & MRIP \\ 
  \midrule
CRC &  0.59 (  0.24,  0.91) &  0.42 (  0.23,  0.61) &  0.41 (  0.23,  0.60) &  0.53 (  0.29,  0.74) &  0.52 (  0.29,  0.73) \\ 
  RCC &  0.44 (-28.08, 34.37) &  0.42 (  0.23,  0.61) &  0.41 (  0.23,  0.60) &  0.43 (  0.14,  0.72) &  0.43 (  0.13,  0.71) \\ 
  BC &  0.40 ( -0.55,  1.60) &  0.42 (  0.23,  0.61) &  0.41 (  0.23,  0.60) &  0.42 (  0.15,  0.70) &  0.41 (  0.14,  0.68) \\ 
  NSCLC &  0.58 (  0.02,  1.12) &  0.42 (  0.23,  0.61) &  0.41 (  0.23,  0.60) &  0.48 (  0.23,  0.70) &  0.47 (  0.23,  0.71) \\ 
  OFTPP &  0.24 ( -0.20,  0.75) &  0.42 (  0.23,  0.61) &  0.41 (  0.23,  0.60) &  0.34 (  0.10,  0.57) &  0.33 (  0.10,  0.57) \\ 
  CC & -0.47 (-54.65, 59.38) &  0.42 (  0.23,  0.61) &  0.41 (  0.22,  0.61) &  0.49 (  0.16,  0.92) &  0.48 (  0.14,  0.93) \\ 
  GBM &  0.30 ( -3.14,  3.47) &  0.42 (  0.23,  0.61) &  0.41 (  0.23,  0.60) &  0.34 (  0.03,  0.60) &  0.33 (  0.02,  0.60) \\ 
  All Indications &  &  0.42 (  0.23,  0.61) &  0.41 (  0.23,  0.60) &  0.43 (  0.20,  0.66) &  0.43 (  0.19,  0.66) \\ 
   \bottomrule
\end{tabular}
\end{table}
\begin{table}[H]
\centering
\caption{Median and 95\% credible interval estimates for the 
conditonal variance parameter in the surrogacy models, where mixture models assume 
                       equal mixture probabilities, using data available at present-day.} 
\label{tab:surrogatePsiPresentEqualMixProb}
\begin{tabular}{lrrrrr}
  \toprule
Indication & IP & CP & MCIP & RP & MRIP \\ 
  \midrule
CRC & 0.05 (0.00, 0.17) & 0.05 (0.00, 0.11) & 0.05 (0.00, 0.11) & 0.04 (0.00, 0.11) & 0.04 (0.00, 0.11) \\ 
  RCC & 0.26 (0.01, 1.02) & 0.05 (0.00, 0.11) & 0.05 (0.00, 0.11) & 0.04 (0.00, 0.15) & 0.04 (0.00, 0.15) \\ 
  BC & 0.07 (0.00, 0.26) & 0.05 (0.00, 0.11) & 0.05 (0.00, 0.11) & 0.04 (0.00, 0.13) & 0.04 (0.00, 0.13) \\ 
  NSCLC & 0.08 (0.00, 0.34) & 0.05 (0.00, 0.11) & 0.05 (0.00, 0.11) & 0.04 (0.00, 0.14) & 0.04 (0.00, 0.14) \\ 
  OFTPP & 0.06 (0.00, 0.30) & 0.05 (0.00, 0.11) & 0.05 (0.00, 0.11) & 0.04 (0.00, 0.14) & 0.04 (0.00, 0.14) \\ 
  CC & 0.34 (0.02, 1.12) & 0.05 (0.00, 0.11) & 0.05 (0.00, 0.11) & 0.05 (0.00, 0.19) & 0.05 (0.00, 0.21) \\ 
  GBM & 0.23 (0.02, 0.91) & 0.05 (0.00, 0.11) & 0.05 (0.00, 0.11) & 0.05 (0.00, 0.18) & 0.05 (0.00, 0.18) \\ 
  All Indications &  & 0.05 (0.00, 0.11) & 0.05 (0.00, 0.11) &  &  \\ 
   \bottomrule
\end{tabular}
\end{table}
			\subsubsection{Deviance information criterion (DIC)} \label{sec:appPresentSurrogateDicEqualMixProb}
\begin{table}[H]
\centering
\caption{Residual deviance, deviance, number of effective parameters, and deviance information criterian (DIC) from the application of surrogacy models, including mixture 
                       models assuming equal mixture weights, to data at present-day. *Based on 38 data points.} 
\label{tab:dicSurrogatePresentEqualMixProb}
\begin{tabular}{lrrrr}
  \toprule
Model & Residual deviance & Deviance & $p_D$ & DIC \\ 
  \midrule
IP & 70.95 & -127.59 & 48.75 & -78.84 \\ 
  CP & 72.75 & -118.81 & 39.52 & -79.29 \\ 
  RP & 72.93 & -109.31 & 39.37 & -69.94 \\ 
  MCIP & 71.92 & -122.36 & 42.80 & -79.56 \\ 
  MRIP & 71.77 & -112.79 & 42.14 & -70.65 \\ 
   \bottomrule
\end{tabular}
\end{table}

			\subsubsection{Mixture probabilities} \label{sec:appPresentSurrogateMixProbsEqualMixProb}
\begin{table}[H]
\centering
\caption{Mixture probability estimates (as mean and standard deviation), and informative and vague mixture component estimates (as median and 95\% credible interval), for the intercept parameter from the application of surrogacy models, 
                       assuming equal mixture probabilities across surrogacy parameters, 
                       to data at present-day. } 
\label{tab:mixProbsInterceptsPresentEqualMixProbWithBranch}
\begin{tabular}{llrrr}
  \toprule
Model & Indication & Probability & Informative & Vague \\ 
  \midrule
MCIP & BC & 1.00 (0.00) & 0.04 (-0.04, 0.12) & -0.06 (-62.35, 62.01) \\ 
  MCIP & CC & 0.99 (0.08) & 0.04 (-0.04, 0.12) &  0.06 (-61.73, 61.76) \\ 
  MCIP & CRC & 1.00 (0.00) & 0.04 (-0.04, 0.12) & -0.01 (-61.93, 61.90) \\ 
  MCIP & GBM & 1.00 (0.03) & 0.04 (-0.04, 0.12) &  0.04 (-62.20, 62.16) \\ 
  MCIP & NSCLC & 1.00 (0.00) & 0.04 (-0.04, 0.12) & -0.01 (-61.82, 62.07) \\ 
  MCIP & OFTPP & 1.00 (0.00) & 0.04 (-0.04, 0.12) &  0.02 (-62.11, 61.97) \\ 
  MCIP & RCC & 1.00 (0.01) & 0.04 (-0.04, 0.12) & -0.21 (-62.35, 61.95) \\ 
 & & & & \\
  MRIP & BC & 1.00 (0.00) & 0.04 (-0.06, 0.14) &  0.07 (-61.95, 61.72) \\ 
  MRIP & CC & 0.99 (0.09) & 0.03 (-0.12, 0.14) &  0.00 (-61.89, 61.92) \\ 
  MRIP & CRC & 1.00 (0.00) & 0.03 (-0.07, 0.12) & -0.06 (-62.06, 61.96) \\ 
  MRIP & GBM & 1.00 (0.01) & 0.06 (-0.04, 0.21) &  0.20 (-61.81, 62.00) \\ 
  MRIP & NSCLC & 1.00 (0.00) & 0.05 (-0.06, 0.15) & -0.02 (-62.13, 62.37) \\ 
  MRIP & OFTPP & 1.00 (0.00) & 0.05 (-0.05, 0.15) &  0.02 (-61.92, 61.87) \\ 
  MRIP & RCC & 1.00 (0.02) & 0.04 (-0.07, 0.16) & -0.04 (-62.16, 62.17) \\ 
   \bottomrule
\end{tabular}
\end{table}
\begin{table}[H]
\centering
\caption{Mixture probability estimates (as mean and standard deviation), and informative and vague mixture component estimates (as median and 95\% credible interval), for the slope parameter from the application of surrogacy models, 
                       assuming equal mixture probabilities across surrogacy parameters, 
                        to data at present-day. } 
\label{tab:mixProbsSlopesPresentEqualMixProbWithBranch}
\begin{tabular}{llrrr}
  \toprule
Model & Indication & Probability & Informative & Vague \\ 
  \midrule
MCIP & BC & 1.00 (0.00) & 0.41 ( 0.23, 0.60) &  0.00 (-61.95, 61.94) \\ 
  MCIP & CC & 0.99 (0.08) & 0.41 ( 0.23, 0.60) & -0.02 (-61.59, 62.38) \\ 
  MCIP & CRC & 1.00 (0.00) & 0.41 ( 0.23, 0.60) & -0.07 (-61.79, 61.92) \\ 
  MCIP & GBM & 1.00 (0.03) & 0.41 ( 0.23, 0.60) &  0.09 (-62.09, 62.04) \\ 
  MCIP & NSCLC & 1.00 (0.00) & 0.41 ( 0.23, 0.60) & -0.10 (-61.96, 62.12) \\ 
  MCIP & OFTPP & 1.00 (0.00) & 0.41 ( 0.23, 0.60) &  0.03 (-61.70, 61.99) \\ 
  MCIP & RCC & 1.00 (0.01) & 0.41 ( 0.23, 0.60) &  0.19 (-61.86, 61.92) \\ 
 & & & & \\
  MRIP & BC & 1.00 (0.00) & 0.41 ( 0.14, 0.68) & -0.14 (-61.80, 61.67) \\ 
  MRIP & CC & 0.99 (0.09) & 0.48 ( 0.15, 0.91) &  0.20 (-61.87, 62.25) \\ 
  MRIP & CRC & 1.00 (0.00) & 0.52 ( 0.29, 0.73) & -0.01 (-61.98, 61.99) \\ 
  MRIP & GBM & 1.00 (0.01) & 0.33 ( 0.02, 0.60) &  0.19 (-61.90, 61.94) \\ 
  MRIP & NSCLC & 1.00 (0.00) & 0.47 ( 0.23, 0.71) & -0.03 (-61.93, 61.92) \\ 
  MRIP & OFTPP & 1.00 (0.00) & 0.33 ( 0.10, 0.57) &  0.04 (-61.99, 62.22) \\ 
  MRIP & RCC & 1.00 (0.02) & 0.43 ( 0.13, 0.71) & -0.05 (-61.92, 61.53) \\ 
   \bottomrule
\end{tabular}
\end{table}
\begin{table}[H]
\centering
\caption{Mixture probability estimates (as mean and standard deviation), and informative and vague mixture component estimates (as median and 95\% credible interval), for conditional variance parameters from the application of surrogacy models, 
                       assuming equal mixture probabilities across surrogacy parameters, 
                        to data at present-day. } 
\label{tab:mixProbsVariancesPresentEqualMixProbWithBranch}
\begin{tabular}{llrrr}
  \toprule
Model & Indication & Probability & Informative & Vague \\ 
  \midrule
MCIP & BC & 1.00 (0.00) & 0.05 ( 0.00, 0.11) &  0.34 (  0.02,  1.11) \\ 
  MCIP & CC & 0.99 (0.08) & 0.05 ( 0.00, 0.11) &  0.34 (  0.02,  1.12) \\ 
  MCIP & CRC & 1.00 (0.00) & 0.05 ( 0.00, 0.11) &  0.34 (  0.02,  1.14) \\ 
  MCIP & GBM & 1.00 (0.03) & 0.05 ( 0.00, 0.11) &  0.34 (  0.02,  1.12) \\ 
  MCIP & NSCLC & 1.00 (0.00) & 0.05 ( 0.00, 0.11) &  0.34 (  0.02,  1.12) \\ 
  MCIP & OFTPP & 1.00 (0.00) & 0.05 ( 0.00, 0.11) &  0.34 (  0.02,  1.12) \\ 
  MCIP & RCC & 1.00 (0.01) & 0.05 ( 0.00, 0.11) &  0.34 (  0.02,  1.12) \\ 
 & & & & \\
  MRIP & BC & 1.00 (0.00) & 0.04 ( 0.00, 0.13) &  0.34 (  0.02,  1.12) \\ 
  MRIP & CC & 0.99 (0.09) & 0.05 ( 0.00, 0.20) &  0.34 (  0.02,  1.12) \\ 
  MRIP & CRC & 1.00 (0.00) & 0.04 ( 0.00, 0.11) &  0.34 (  0.02,  1.12) \\ 
  MRIP & GBM & 1.00 (0.01) & 0.05 ( 0.00, 0.18) &  0.34 (  0.02,  1.12) \\ 
  MRIP & NSCLC & 1.00 (0.00) & 0.04 ( 0.00, 0.14) &  0.34 (  0.02,  1.12) \\ 
  MRIP & OFTPP & 1.00 (0.00) & 0.04 ( 0.00, 0.14) &  0.34 (  0.02,  1.12) \\ 
  MRIP & RCC & 1.00 (0.02) & 0.04 ( 0.00, 0.15) &  0.34 (  0.02,  1.12) \\ 
   \bottomrule
\end{tabular}
\end{table}
                \clearpage
             \end{landscape}

            \subsection{Sensitivity analysis: independent conditional variance parameters}
                \subsubsection{Surrogacy parameter estimates} \label{sec:appPresentSurrogacyIndependentPsi}
                \begin{figure}[!htb]
				\centering
				\includegraphics[width=\textwidth]{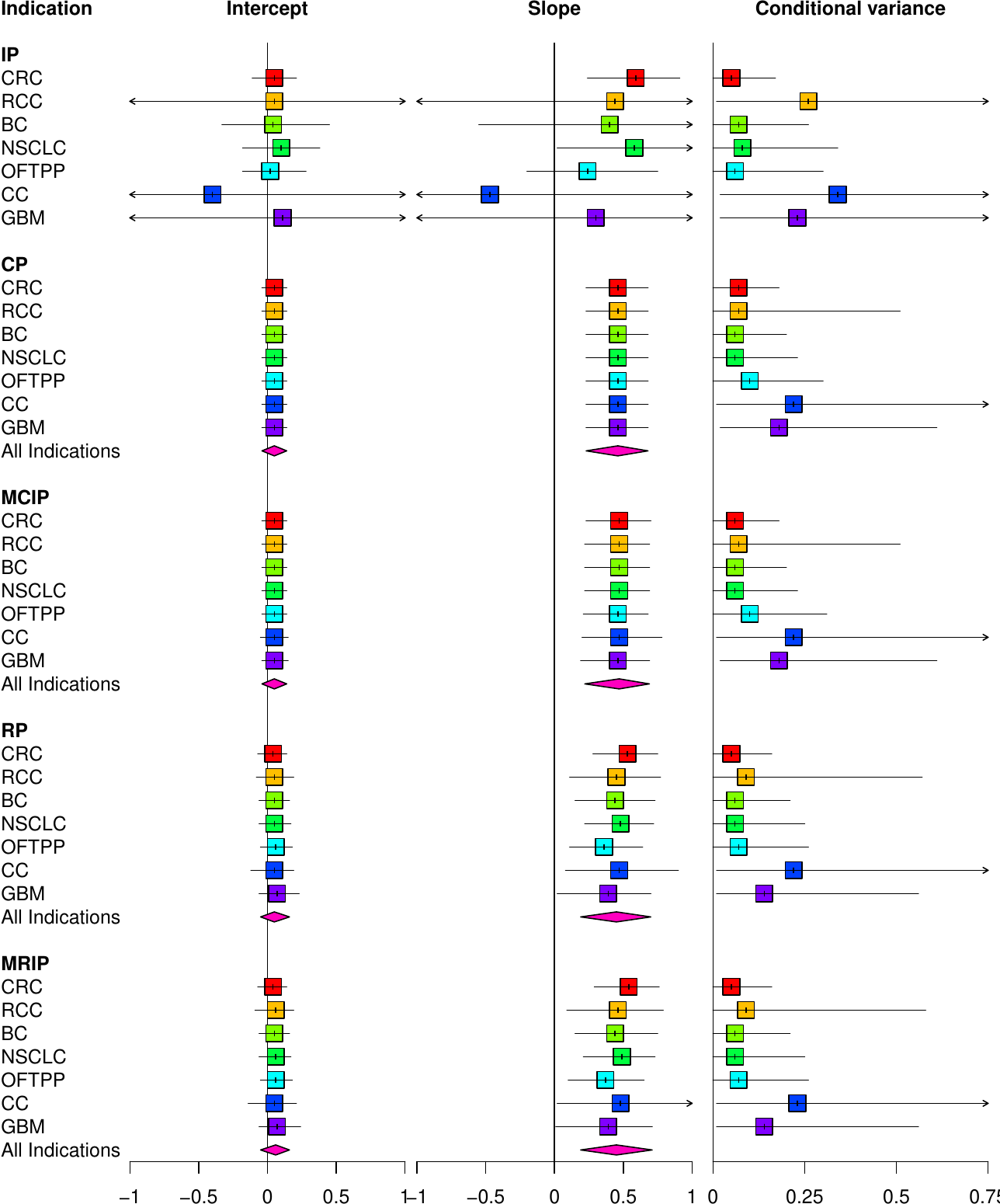}
				\caption{Multi-indication meta-analysis of surrogacy parameters, assuming independent conditional variance parameters, using data available at present day. IP - independent parameters, CP - common parameter, MCIP - mixed common and independent parameters, RP - random parameters, MRIP - mixed random and independent parameters, CRC - colorectal cancer, BC - breast cancer, NSCLC - non-small cell lung cancer, RCC - renal cell carcinoma, CC - cervical cancer, GBM - glioblastoma, OFTPP - ovarian, fallopian tube, and primary peritoneal.}
				\label{fig:presentDaySurrogacyIndependentPsi}
			\end{figure}
			\clearpage

             \begin{landscape}
\begin{table}[H]
\centering
\caption{Median and 95\% credible interval estimates for the intercept parameter
                       in the surrogacy models, assuming independent conditional variance parameters,
                       using data available at present-day.} 
\label{tab:surrogateInterceptPresentIndependentPsi}
\begin{tabular}{lrrrrr}
  \toprule
Indication & IP & CP & MCIP & RP & MRIP \\ 
  \midrule
CRC &  0.05 ( -0.11,  0.21) &  0.05 ( -0.04,  0.14) &  0.05 ( -0.04,  0.14) &  0.04 ( -0.07,  0.14) &  0.04 ( -0.07,  0.14) \\ 
  RCC &  0.05 (-12.75, 15.19) &  0.05 ( -0.04,  0.14) &  0.05 ( -0.04,  0.14) &  0.05 ( -0.08,  0.19) &  0.06 ( -0.09,  0.19) \\ 
  BC &  0.04 ( -0.33,  0.45) &  0.05 ( -0.04,  0.14) &  0.05 ( -0.04,  0.14) &  0.05 ( -0.06,  0.16) &  0.05 ( -0.06,  0.16) \\ 
  NSCLC &  0.10 ( -0.18,  0.38) &  0.05 ( -0.04,  0.14) &  0.05 ( -0.04,  0.14) &  0.05 ( -0.06,  0.17) &  0.06 ( -0.06,  0.17) \\ 
  OFTPP &  0.02 ( -0.18,  0.28) &  0.05 ( -0.04,  0.14) &  0.05 ( -0.04,  0.14) &  0.06 ( -0.05,  0.18) &  0.06 ( -0.05,  0.18) \\ 
  CC & -0.40 (-22.89, 23.86) &  0.05 ( -0.04,  0.14) &  0.05 ( -0.05,  0.15) &  0.05 ( -0.12,  0.19) &  0.05 ( -0.14,  0.21) \\ 
  GBM &  0.11 ( -1.52,  1.64) &  0.05 ( -0.04,  0.14) &  0.05 ( -0.04,  0.15) &  0.07 ( -0.06,  0.23) &  0.07 ( -0.06,  0.24) \\ 
  All Indications &  &  0.05 ( -0.04,  0.14) &  0.05 ( -0.04,  0.14) &  0.05 ( -0.05,  0.16) &  0.06 ( -0.05,  0.16) \\ 
   \bottomrule
\end{tabular}
\end{table}
\begin{table}[H]
\centering
\caption{Median and 95\% credible interval estimates for the slope parameter
                       in the surrogacy models assuming independent conditional variance parameters,
                       using data available at present-day.} 
\label{tab:surrogateSlopePresentIndependentPsi}
\begin{tabular}{lrrrrr}
  \toprule
Indication & IP & CP & MCIP & RP & MRIP \\ 
  \midrule
CRC &  0.59 (  0.24,  0.91) &  0.46 (  0.23,  0.68) &  0.47 (  0.23,  0.70) &  0.53 (  0.28,  0.75) &  0.54 (  0.29,  0.76) \\ 
  RCC &  0.44 (-28.08, 34.37) &  0.46 (  0.23,  0.68) &  0.47 (  0.22,  0.69) &  0.45 (  0.11,  0.77) &  0.46 (  0.09,  0.79) \\ 
  BC &  0.40 ( -0.55,  1.60) &  0.46 (  0.23,  0.68) &  0.47 (  0.22,  0.69) &  0.44 (  0.15,  0.73) &  0.44 (  0.15,  0.75) \\ 
  NSCLC &  0.58 (  0.02,  1.12) &  0.46 (  0.23,  0.68) &  0.47 (  0.22,  0.69) &  0.48 (  0.22,  0.72) &  0.49 (  0.21,  0.73) \\ 
  OFTPP &  0.24 ( -0.20,  0.75) &  0.46 (  0.23,  0.68) &  0.46 (  0.21,  0.68) &  0.36 (  0.11,  0.64) &  0.37 (  0.10,  0.65) \\ 
  CC & -0.47 (-54.65, 59.38) &  0.46 (  0.23,  0.68) &  0.47 (  0.20,  0.78) &  0.47 (  0.08,  0.90) &  0.48 (  0.02,  1.11) \\ 
  GBM &  0.30 ( -3.14,  3.47) &  0.46 (  0.23,  0.68) &  0.46 (  0.19,  0.69) &  0.39 (  0.02,  0.70) &  0.39 (  0.01,  0.71) \\ 
  All Indications &  &  0.46 (  0.23,  0.68) &  0.47 (  0.22,  0.69) &  0.45 (  0.19,  0.70) &  0.45 (  0.19,  0.71) \\ 
   \bottomrule
\end{tabular}
\end{table}
\begin{table}[H]
\centering
\caption{Median and 95\% credible interval estimates for the 
conditonal variance parameter in the surrogacy models assuming independent conditional variance parameters,
                       using data available at present-day.} 
\label{tab:surrogatePsiPresentIndependentPsi}
\begin{tabular}{lrrrrr}
  \toprule
Indication & IP & CP & MCIP & RP & MRIP \\ 
  \midrule
CRC & 0.05 (0.00, 0.17) & 0.07 (0.00, 0.18) & 0.06 (0.00, 0.18) & 0.05 (0.00, 0.16) & 0.05 (0.00, 0.16) \\ 
  RCC & 0.26 (0.01, 1.02) & 0.07 (0.00, 0.51) & 0.07 (0.00, 0.51) & 0.09 (0.00, 0.57) & 0.09 (0.00, 0.58) \\ 
  BC & 0.07 (0.00, 0.26) & 0.06 (0.00, 0.20) & 0.06 (0.00, 0.20) & 0.06 (0.00, 0.21) & 0.06 (0.00, 0.21) \\ 
  NSCLC & 0.08 (0.00, 0.34) & 0.06 (0.00, 0.23) & 0.06 (0.00, 0.23) & 0.06 (0.00, 0.25) & 0.06 (0.00, 0.25) \\ 
  OFTPP & 0.06 (0.00, 0.30) & 0.10 (0.00, 0.30) & 0.10 (0.00, 0.31) & 0.07 (0.00, 0.26) & 0.07 (0.00, 0.26) \\ 
  CC & 0.34 (0.02, 1.12) & 0.22 (0.01, 0.90) & 0.22 (0.01, 0.92) & 0.22 (0.01, 0.91) & 0.23 (0.01, 0.92) \\ 
  GBM & 0.23 (0.02, 0.91) & 0.18 (0.02, 0.61) & 0.18 (0.02, 0.61) & 0.14 (0.01, 0.56) & 0.14 (0.01, 0.56) \\ 
   \bottomrule
\end{tabular}
\end{table}
			\subsubsection{Deviance information criterion (DIC)} \label{sec:appPresentSurrogateDicIndependentPsi}
\begin{table}[H]
\centering
\caption{Residual deviance, number of effective parameters, and deviance information criterian (DIC) from the application of surrogacy models assuming independent conditional variance parameters
                       to data at present-day. *Based on 38 data points.} 
\label{tab:dicSurrogatePresentIndependentPsi}
\begin{tabular}{lrrrr}
  \toprule
Model & Residual deviance & Deviance & $p_D$ & DIC \\ 
  \midrule
IP & 70.95 & -127.59 & 48.75 & -78.84 \\ 
  CP & 71.62 & -125.80 & 44.36 & -81.44 \\ 
  MCIP & 71.49 & -125.88 & 45.01 & -80.86 \\ 
  RP & 71.36 & -128.82 & 45.04 & -83.79 \\ 
  MRIP & 71.41 & -128.01 & 45.36 & -82.65 \\ 
   \bottomrule
\end{tabular}
\end{table}

			\subsubsection{Mixture probabilities} \label{sec:appPresentSurrogateMixProbsIndependentPsi}
\begin{table}[H]
\centering
\caption{Mixture probability estimates (as mean and standard deviation), and informative and vague mixture component estimates (as median and 95\% credible interval), for the intercept parameter from the application of surrogacy models 
                       assuming independent conditional variance parameters 
                       to data at present-day. } 
\label{tab:mixProbsInterceptsPresentIndependentPsiWithBranch}
\begin{tabular}{llrrr}
  \toprule
Model & Indication & Probability & Informative & Vague \\ 
  \midrule
MCIP & BC & 1.00 (0.03) & 0.05 (-0.04, 0.14) &  0.09 (-61.55, 62.29) \\ 
  MCIP & CC & 0.98 (0.14) & 0.05 (-0.04, 0.14) &  0.03 (-61.70, 61.62) \\ 
  MCIP & CRC & 0.99 (0.08) & 0.05 (-0.04, 0.14) & -0.01 (-62.29, 61.80) \\ 
  MCIP & GBM & 0.99 (0.10) & 0.05 (-0.04, 0.14) &  0.15 (-62.04, 61.75) \\ 
  MCIP & NSCLC & 1.00 (0.05) & 0.05 (-0.04, 0.14) &  0.03 (-62.26, 62.03) \\ 
  MCIP & OFTPP & 1.00 (0.07) & 0.05 (-0.04, 0.14) &  0.09 (-61.54, 62.38) \\ 
  MCIP & RCC & 1.00 (0.06) & 0.05 (-0.04, 0.14) & -0.03 (-61.98, 62.25) \\ 
 & & & & \\
  MRIP & BC & 1.00 (0.05) & 0.05 (-0.06, 0.16) & -0.19 (-62.06, 61.65) \\ 
  MRIP & CC & 0.98 (0.14) & 0.05 (-0.12, 0.20) &  0.05 (-61.73, 61.71) \\ 
  MRIP & CRC & 1.00 (0.07) & 0.04 (-0.07, 0.14) &  0.05 (-61.91, 61.80) \\ 
  MRIP & GBM & 0.99 (0.09) & 0.07 (-0.06, 0.23) &  0.13 (-61.92, 61.71) \\ 
  MRIP & NSCLC & 1.00 (0.05) & 0.06 (-0.06, 0.17) &  0.06 (-62.08, 61.94) \\ 
  MRIP & OFTPP & 1.00 (0.06) & 0.06 (-0.05, 0.18) &  0.01 (-61.93, 62.03) \\ 
  MRIP & RCC & 0.99 (0.09) & 0.06 (-0.09, 0.19) & -0.04 (-61.95, 61.60) \\ 
   \bottomrule
\end{tabular}
\end{table}
\begin{table}[H]
\centering
\caption{Mixture probability estimates (as mean and standard deviation), and informative and vague mixture component estimates (as median and 95\% credible interval), for the slope parameter from the application of surrogacy models 
                       assuming independent conditional variance parameters 
                        to data at present-day. } 
\label{tab:mixProbsSlopesPresentIndependentPsiWithBranch}
\begin{tabular}{llrrr}
  \toprule
Model & Indication & Probability & Informative & Vague \\ 
  \midrule
MCIP & BC & 1.00 (0.06) & 0.47 ( 0.22, 0.69) &  0.22 (-61.85, 61.96) \\ 
  MCIP & CC & 0.96 (0.20) & 0.47 ( 0.22, 0.69) &  0.55 (-61.49, 61.38) \\ 
  MCIP & CRC & 0.97 (0.17) & 0.47 ( 0.22, 0.69) &  0.52 (-61.69, 61.95) \\ 
  MCIP & GBM & 0.97 (0.16) & 0.47 ( 0.22, 0.69) &  0.16 (-61.87, 61.65) \\ 
  MCIP & NSCLC & 0.99 (0.07) & 0.47 ( 0.22, 0.69) &  0.21 (-61.93, 61.86) \\ 
  MCIP & OFTPP & 0.97 (0.16) & 0.47 ( 0.22, 0.69) &  0.26 (-61.49, 61.18) \\ 
  MCIP & RCC & 0.99 (0.09) & 0.47 ( 0.22, 0.69) &  0.21 (-61.89, 61.97) \\ 
 & & & & \\
  MRIP & BC & 0.99 (0.09) & 0.44 ( 0.15, 0.75) &  0.23 (-61.95, 61.83) \\ 
  MRIP & CC & 0.96 (0.20) & 0.48 ( 0.07, 0.92) &  0.52 (-61.27, 61.66) \\ 
  MRIP & CRC & 0.99 (0.11) & 0.54 ( 0.28, 0.76) &  0.37 (-62.10, 61.72) \\ 
  MRIP & GBM & 0.98 (0.13) & 0.39 ( 0.02, 0.71) &  0.13 (-61.26, 61.56) \\ 
  MRIP & NSCLC & 0.99 (0.08) & 0.49 ( 0.21, 0.73) &  0.22 (-62.03, 61.99) \\ 
  MRIP & OFTPP & 0.99 (0.11) & 0.37 ( 0.10, 0.66) &  0.23 (-61.87, 61.56) \\ 
  MRIP & RCC & 0.99 (0.12) & 0.46 ( 0.10, 0.78) &  0.14 (-61.95, 61.78) \\ 
   \bottomrule
\end{tabular}
\end{table}
                \clearpage
             \end{landscape}

            \subsection{Sensitivity analysis: excluding data on breast cancer}
                \subsubsection{Surrogacy parameter estimates} \label{sec:appPresentSurrogacyBreast}
                \begin{figure}[!htb]
				\centering
				\includegraphics[width=\textwidth]{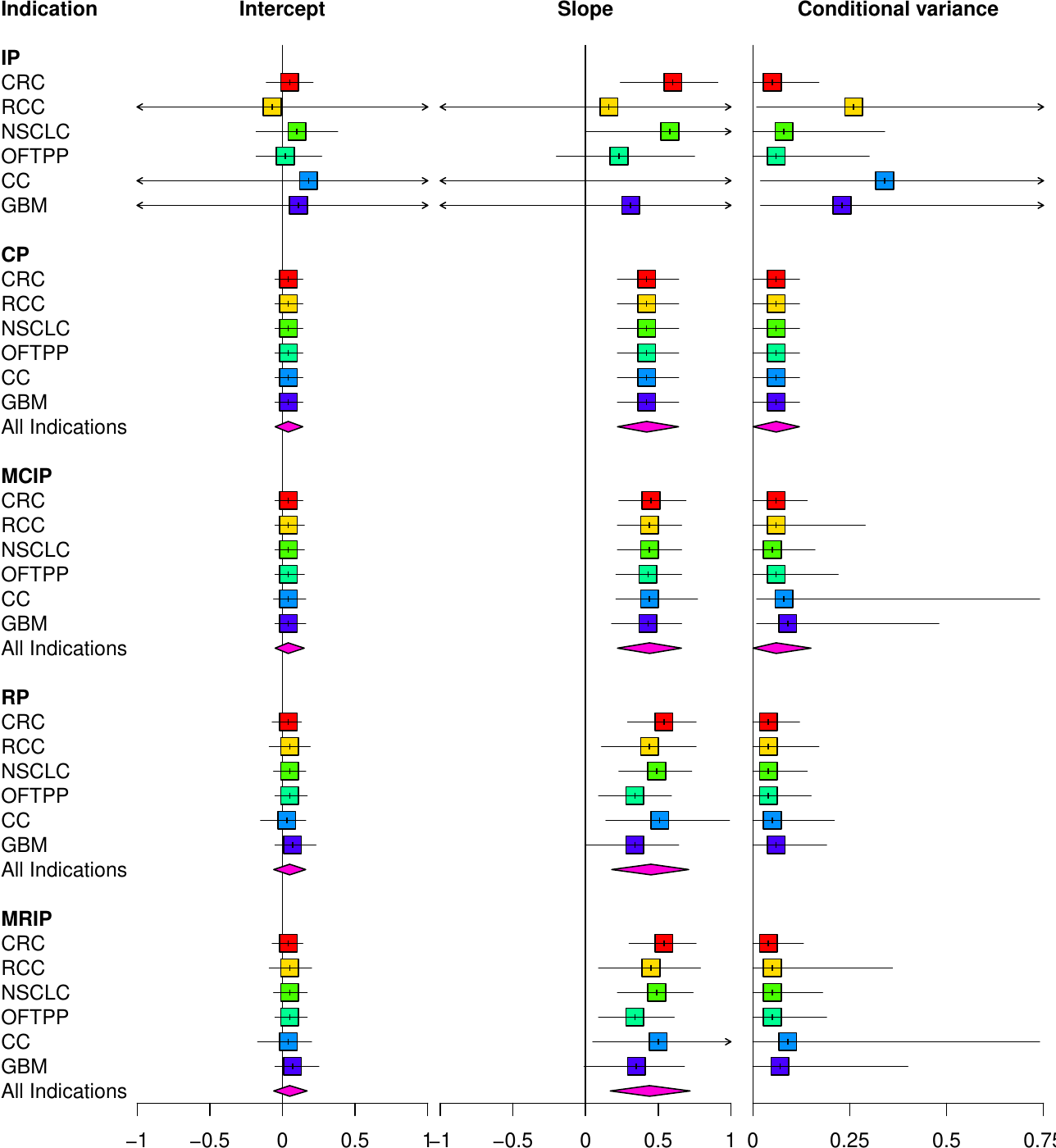}
				\caption{Multi-indication meta-analysis of surrogacy parameters using data available at present day, excluding data from trials in breast cancer. IP - independent parameters, CP - common parameter, MCIP - mixed common and independent parameters, RP - random parameters, MRIP - mixed random and independent parameters, CRC - colorectal cancer, BC - breast cancer, NSCLC - non-small cell lung cancer, RCC - renal cell carcinoma, CC - cervical cancer, GBM - glioblastoma, OFTPP - ovarian, fallopian tube, and primary peritoneal.}
				\label{fig:presentDaySurrogacyBreast}
			\end{figure}
			\clearpage

             \begin{landscape}
\begin{table}[H]
\centering
\caption{Median and 95\% credible interval estimates for the intercept parameter
                       in the surrogacy models,
                       using data available at present-day excluding breast cancer.} 
\label{tab:surrogateInterceptPresentBreast}
\begin{tabular}{lrrrrr}
  \toprule
Indication & IP & CP & MCIP & RP & MRIP \\ 
  \midrule
CRC &  0.05 ( -0.11,  0.21) &  0.04 ( -0.05,  0.14) &  0.04 ( -0.05,  0.14) &  0.04 ( -0.07,  0.13) &  0.04 ( -0.07,  0.14) \\ 
  RCC & -0.07 (-15.57, 14.90) &  0.04 ( -0.05,  0.14) &  0.04 ( -0.05,  0.15) &  0.05 ( -0.09,  0.19) &  0.05 ( -0.09,  0.20) \\ 
  NSCLC &  0.10 ( -0.18,  0.38) &  0.04 ( -0.05,  0.14) &  0.04 ( -0.05,  0.15) &  0.05 ( -0.06,  0.16) &  0.05 ( -0.06,  0.17) \\ 
  OFTPP &  0.02 ( -0.18,  0.27) &  0.04 ( -0.05,  0.14) &  0.04 ( -0.05,  0.15) &  0.05 ( -0.05,  0.17) &  0.05 ( -0.05,  0.17) \\ 
  CC &  0.18 (-22.65, 23.18) &  0.04 ( -0.05,  0.14) &  0.04 ( -0.06,  0.16) &  0.03 ( -0.15,  0.16) &  0.04 ( -0.17,  0.20) \\ 
  GBM &  0.11 ( -1.59,  1.79) &  0.04 ( -0.05,  0.14) &  0.04 ( -0.05,  0.16) &  0.07 ( -0.05,  0.23) &  0.07 ( -0.05,  0.25) \\ 
  All Indications &  &  0.04 ( -0.05,  0.14) &  0.04 ( -0.05,  0.15) &  0.05 ( -0.06,  0.16) &  0.05 ( -0.06,  0.17) \\ 
   \bottomrule
\end{tabular}
\end{table}
			\input{"./tables/surrogateSlopePresentBreast.txt"}
                \input{"./tables/surrogatePsiPresentBreast.txt"}
			\subsubsection{Deviance information criterion (DIC)} \label{sec:appPresentSurrogateDicBreast}
\begin{table}[H]
\centering
\caption{Residual deviance, number of effective parameters, and deviance information criterian (DIC) from the application of surrogacy models
                       to data at present-day excluding breast cancer. *Based on 28 data points.} 
\label{tab:dicSurrogatePresentBreast}
\begin{tabular}{lrrrr}
  \toprule
Model & Residual deviance & Deviance & $p_D$ & DIC \\ 
  \midrule
IP & 51.28 & -99.09 & 36.04 & -63.05 \\ 
  CP & 52.71 & -93.98 & 30.04 & -63.94 \\ 
  MCIP & 52.37 & -90.34 & 31.71 & -58.63 \\ 
  RP & 51.43 & -96.24 & 32.16 & -64.08 \\ 
  MRIP & 51.02 & -88.42 & 32.97 & -55.45 \\ 
   \bottomrule
\end{tabular}
\end{table}

			\subsubsection{Mixture probabilities} \label{sec:appPresentSurrogateMixProbsBreast}
			\input{"./tables/mixProbsInterceptsPresentBreastWithBranch.txt"}
			\input{"./tables/mixProbsSlopesPresentBreastWithBranch.txt"}
                \clearpage
             \end{landscape}

            \subsection{Sensitivity analysis: common effect across trials within each indication}
                \subsubsection{Treatment effect estimates} \label{sec:appPresentCommonEffect}
                \begin{figure}[!htb]
				\centering
				\includegraphics[height=\textheight]{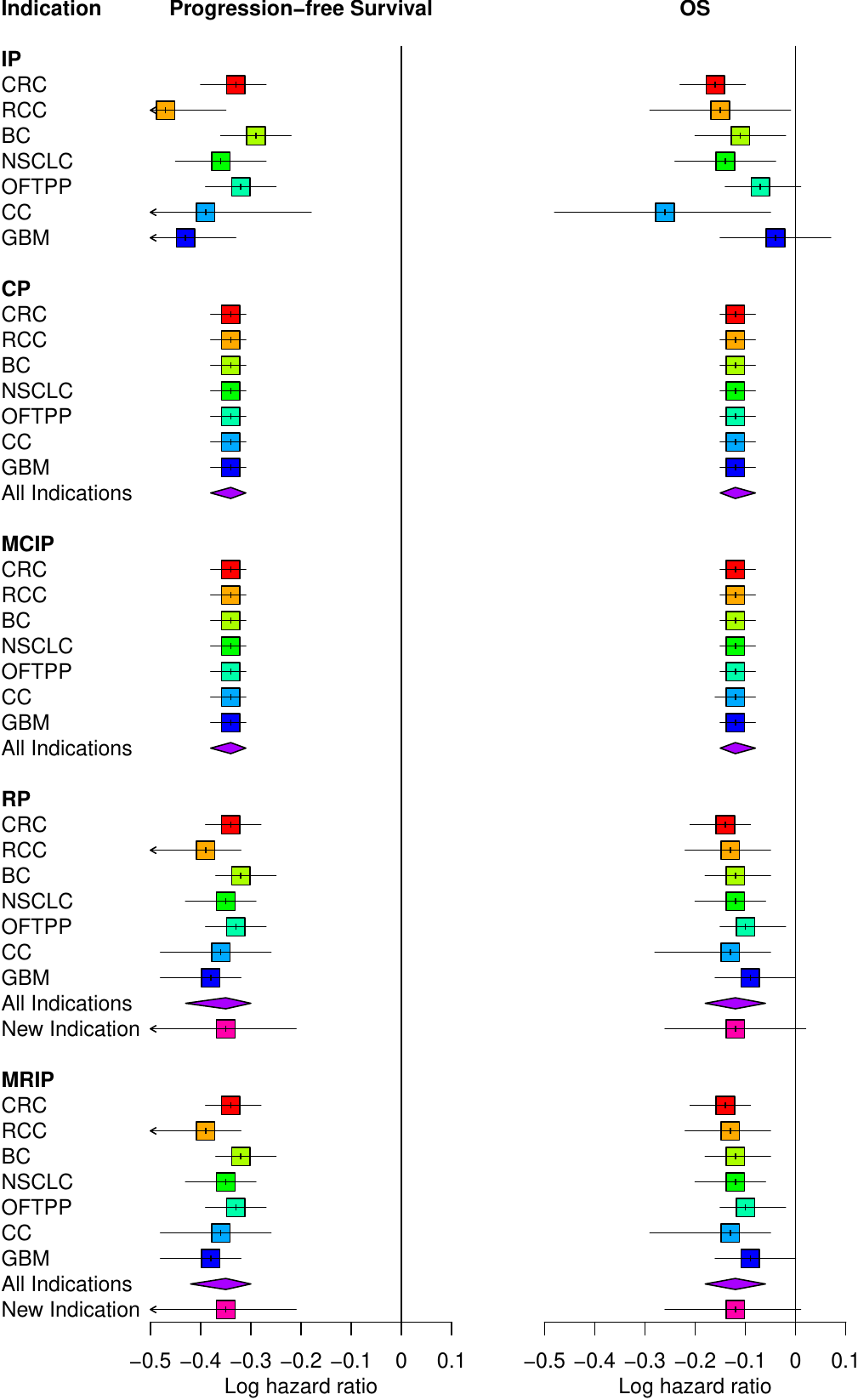}
				\caption{Multi-indication meta-analysis, assuming a common effect across trials within each indication, synthesising effects on overall survival (OS) and progression-free survival (PFS) using data available at present day. IP - independent parameters, CP - common parameter, MCIP - mixed common and independent parameters, RP - random parameters, MRIP - mixed random and independent parameters, CRC - colorectal cancer, BC - breast cancer, NSCLC - non-small cell lung cancer, RCC - renal cell carcinoma, CC - cervical cancer, GBM - glioblastoma, OFTPP - ovarian, fallopian tube, and primary peritoneal.}
				\label{fig:presentDayUnivariateCommonEffect}
			\end{figure}
			\clearpage

             \begin{landscape}
                \input{"./tables/univariateOsPresentCommonEffect.txt"}
			\input{"./tables/univariatePfsPresentCommonEffect.txt"}
			\subsubsection{Deviance information criterion (DIC)} \label{sec:appPresentUnivariateDicCommon}
\begin{table}[H]
\centering
\caption{Residual deviance, deviance, number of effective parameters, and deviance information criterian (DIC) from the application of univariate models, assuming a common effect within indications, to data on overall survival at present-day. *Based on 38 data points.} 
\label{tab:dicCommonOsPresent}
\begin{tabular}{lrrrr}
  \toprule
Model & Residual deviance & Deviance & $p_D$ & DIC \\ 
  \midrule
IP & 48.25 & -34.07 & 7.01 & -27.06 \\ 
  CP & 50.19 & -32.13 & 1.01 & -31.12 \\ 
  MCIP & 50.17 & -32.16 & 1.05 & -31.10 \\ 
  RP & 48.14 & -34.19 & 3.75 & -30.44 \\ 
  MRIP & 48.13 & -32.81 & 4.05 & -28.76 \\ 
   \bottomrule
\end{tabular}
\end{table}

\begin{table}[H]
\centering
\caption{Residual deviance, deviance, number of effective parameters, and deviance information criterian (DIC) from the application of univariate models, assuming a common effect within indications, to data on progression-free survival at present-day. *Based on 43 data points.} 
\label{tab:dicCommonPfsPresent}
\begin{tabular}{lrrrr}
  \toprule
Model & Residual deviance & Deviance & $p_D$ & DIC \\ 
  \midrule
IP & 147.07 & 50.13 & 7.01 & 57.14 \\ 
  CP & 151.29 & 54.34 & 1.01 & 55.35 \\ 
  MCIP & 151.21 & 54.26 & 1.17 & 55.43 \\ 
  RP & 147.42 & 50.47 & 4.53 & 55.00 \\ 
  MRIP & 147.42 & 51.86 & 5.80 & 57.67 \\ 
   \bottomrule
\end{tabular}
\end{table}

			\subsubsection{Mixture probabilities} \label{sec:appPresentUnivariateMixProbsCommon}
			\input{"./tables/mixProbsOsPresentCommonEffectWithBranch.txt"}
			\input{"./tables/mixProbsPfsPresentCommonEffectWithBranch.txt"}
                \clearpage
             \end{landscape}

	\section{Abbreviations}
        \setcounter{table}{0}
	\renewcommand{\thetable}{B\arabic{table}}
	
            \subsection{Treatment names} \label{sec:treatmentNames}
                \input{"./tables/treatmentAbbreviations.txt"}
                \clearpage

        \section{Appraisal time point data sets} \label{sec:appraisalDataSets}
        \setcounter{figure}{0}
	\renewcommand\thefigure{\thesection\arabic{figure}}
	
            \subsection{NICE TA178}
			\begin{sidewaysfigure}[ht]
				\centering
				\includegraphics[width=\textwidth]{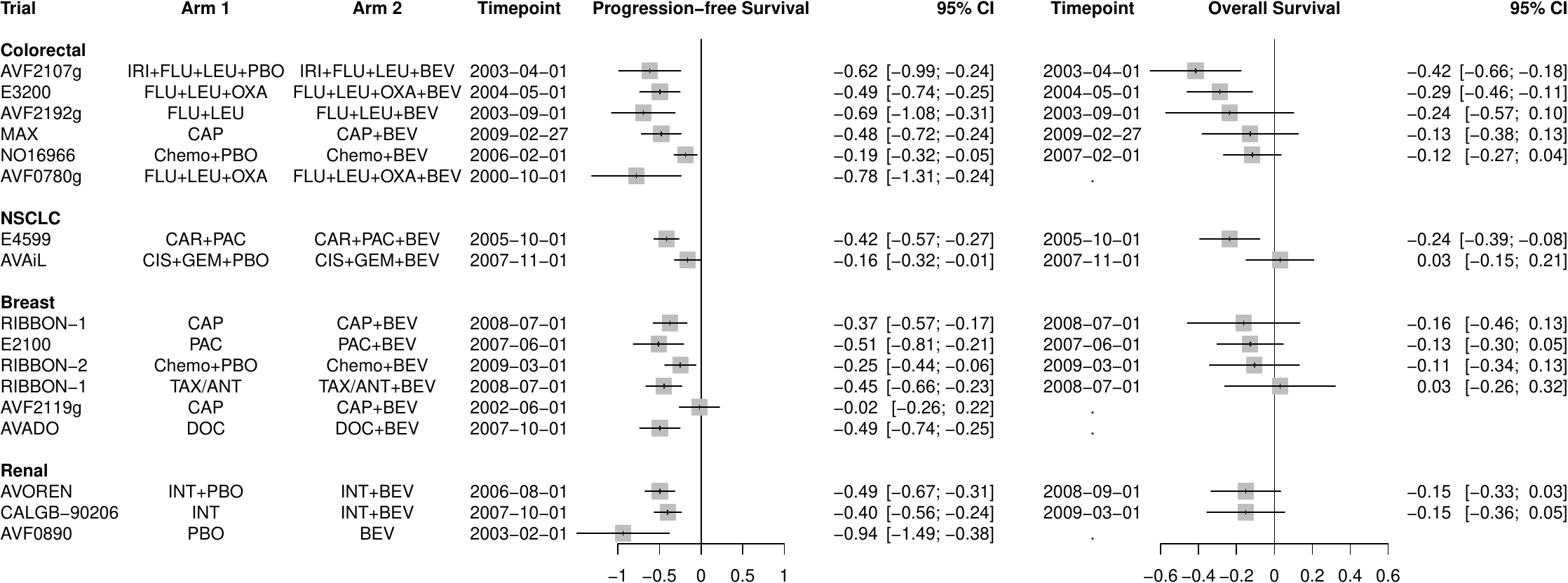}
				\caption{Forest plot summarising log-transformed hazard ratio estimates for effects on progression-free survival and overall survival from randomised controlled trials assessing bevacizumab across cancer indications, for the data available at the time of NICE TA178. Estimates are ordered by OS effect size within each cancer indication. BEV - bevacizumab. NSCLC - Non-small cell lung cancer. OFTPP - Ovarian, fallopian tube and primary peritoneal.}
				\label{fig:observedTa178}
			\end{sidewaysfigure}
			\clearpage

            \subsection{NICE TA285}
			\begin{sidewaysfigure}[ht]
				\centering
				\includegraphics[width=\textwidth]{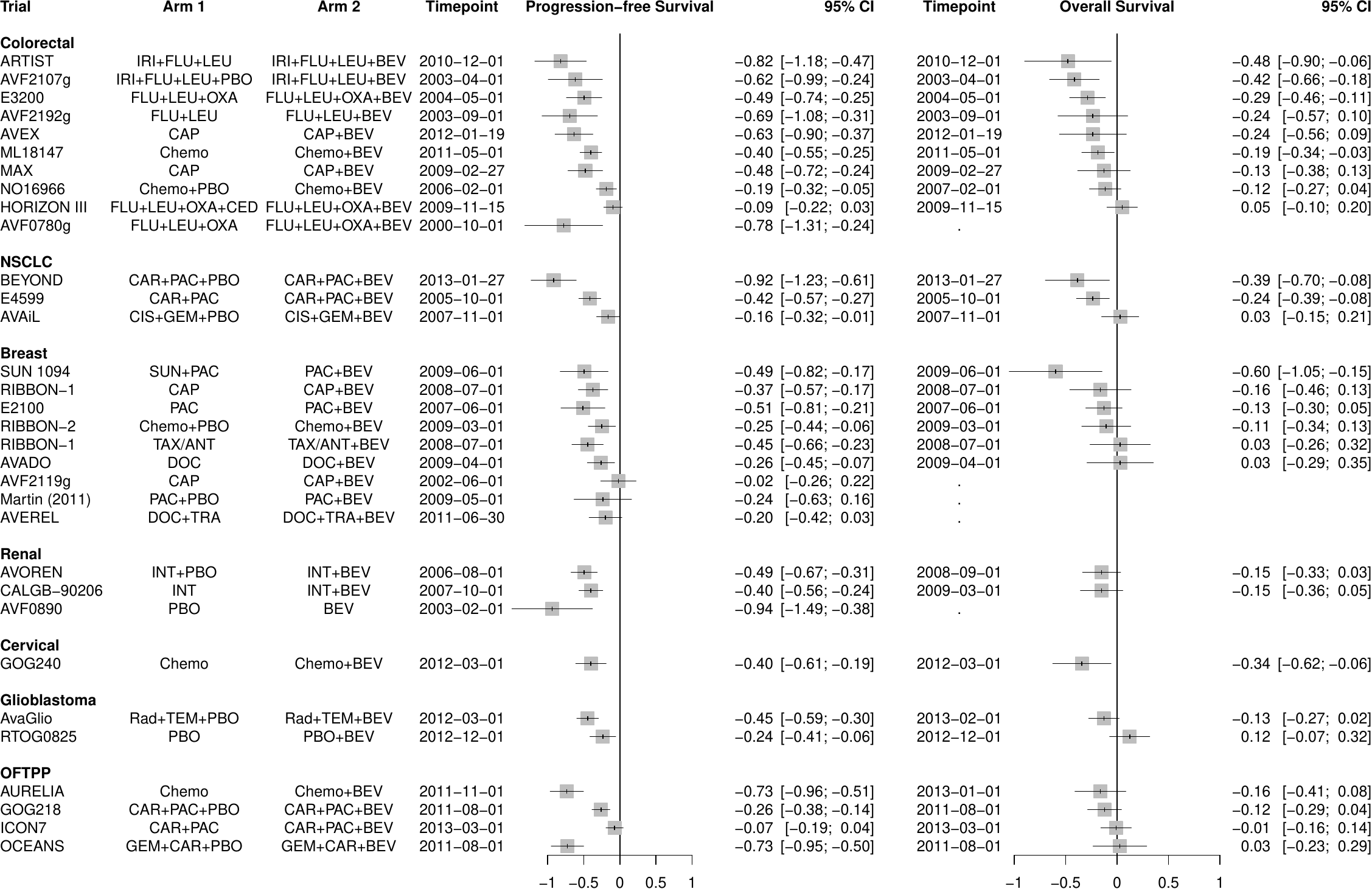}
				\caption{Forest plot summarising log-transformed hazard ratio estimates for effects on progression-free survival and overall survival from randomised controlled trials assessing bevacizumab across cancer indications, for the data available at the time of NICE TA285. Estimates are ordered by OS effect size within each cancer indication. BEV - bevacizumab. NSCLC - Non-small cell lung cancer. OFTPP - Ovarian, fallopian tube and primary peritoneal.}
				\label{fig:observedTa285}
			\end{sidewaysfigure}
			\clearpage

        \section{Formulae for predicting estimates of a treatment effect on overall survival} \label{sec:predictionFormulae}
            A predicted treatment effect estimate on overall survival (OS) can be obtained by entering the indication-specific progression-free survival (PFS) effect estimate into the model describing the surrogate relationship for a particular indication $j$.
            To account for uncertainty in the PFS estimate, the posterior mean $\hat{d}_{j,PFS}$ and posterior standard deviation $\hat{\sigma}_{j,PFS}^2$ are used to obtain a sampled value,
            \begin{equation}
				d_{j, PFS} \sim N(\hat{d}_{j, PFS}, \hat{\sigma}_{d_{j, PFS}}^2),
		\end{equation}
            where $d_{j, PFS}$ represents the sampled PFS estimate.

            The sampled PFS estimate is entered into the regression equation describing the surrogate relationship (i.e. intercept and slope) in indication $j$,
            \begin{equation}
				d_{j, OS} = \lambda_{0j} + \lambda_{1j}d_{j, PFS}
		\end{equation}
            where $d_{j, OS}$ is the predicted treatment effect on OS.
\end{document}